\documentclass{ws-ijmpa}
\usepackage{amssymb,amsmath,stmaryrd}
\usepackage{cite}

\newcommand{\eqna}[1]{\begin{subequations} \label{#1}
\begin{eqnarray}}
\def\eena{\end{eqnarray}
\end{subequations}}

\def\beq{\begin{equation}}
\def\eqn#1{\begin{equation}\label{#1}}
\def\eeq{\end{equation}}

\def\bea{\begin{eqnarray}}
\def\eqnn#1{\bea\label{#1}}
\def\eea{\end{eqnarray}}

\renewcommand{\theequation}{\arabic{section}.\arabic{equation}}

\def\salg{{Schr\"odinger} algebra}
\def\sgl{{Schr\"odinger} equation}

\def\verma{{Verma} module}
\def\mt{\mapsto}
\def\trhalf{{\textstyle{3\over2}}}

\def\rg{\rangle}
\def\cq{{\cal Q}}
\def\hpt{P_t}
\def\hk{K}
\def\hx{P_x}
\def\hg{G}
\def\z{\zeta}
\def\ha{{\textstyle{1\over2}}}
\newcommand{\al}{\alpha}
\newcommand{\be}{\beta}
\newcommand{\II}{{\rm i}}
\def\hp{{\varphi}}
\def\ch{{\cal H}}\def\cd{{\cal D}}
\def\L{{\Lambda}}
\def\l{{\lambda}}
\def\hv{{\hat v}}
\def\vr{\vert}
\def\rg{\rangle}
\def\tu{{\tilde u}}
\def\lra{\longrightarrow}
\def\dia{\diamondsuit}
\def\half{{\textstyle{1\over2}}}
\def\pd{\partial}
\def\cl{{\cal L}}
\def\S{Schr\"odinger$\,$ algebra}
\def\SG{Schr\"odinger$\,$ group}
\def\bu{$\bullet$~}
\def\nt{\noindent}
\def\om{\omega}
\def\eqreff#1#2{(\ref{#1}#2)}
\def\reff#1#2{\ref{#1}#2}

\def\cn{{\cal N}} \def\cm{{\cal M}}

\font\verysmall=cmr13

\def\phr{{\verysmall +}\kern-9pt\supset}

\def\bhr{\raise1.1pt\hbox{\verysmall x}\kern-9pt\subset}
\def\nn{\nonumber}

\def\sch#1{\hat{\cal{S}}(#1)}

\def\cs{{\cal S}}
\def\hs{{\hat \cs}}

\newcommand{\sae}{$\hs$} %

\def\sc{\hat{\cal{S}}}
\def\del#1{\frac{\partial}{\partial #1}}
\def\D{{\Delta}} \def\d{{\delta}}
\def\hX{{\hat X}}
\def\tX{{\tilde X}}
\def\tL{{\tilde L}}
\def\a{{\alpha}}  \def\cb{{\cal B}}\def\cc{{\cal C}}
\def\ce{{\cal E}} \def\te{{\tilde {\cal E}}}
\def\cw{{\cal W}} \def\tw{{\tilde {\cal W}}}
\def\cg{{\cal G}} \def\hhg{{\hat \cg}}
\def\tg{{\tilde {\cal G}}}

\def\np{\newpage}
\def\np{{}}
\def\G{\Gamma}
\def\b{\beta}

\newcommand{\bbr}{\mathbb{R}}
\newcommand{\bbc}{\mathbb{C}}
\newcommand{\bbz}{\mathbb{Z}}
\newcommand{\bbn}{\mathbb{N}}

\begin{document}

\title{{\large Non-Relativistic Holography}\\
-- A Group-Theoretical Perspective}

\author{V.K. Dobrev}

 \address{Institute for Nuclear Research and Nuclear Energy,\\
 Bulgarian Academy of Sciences,\\ 72
Tsarigradsko Chaussee, 1784 Sofia, Bulgaria\\
$^*$E-mail: dobrev@inrne.bas.bg}

\maketitle

\begin{abstract}
We give a review of some group-theoretical results related to non-relativistic holography.
Our  main playgrounds are the Schr\"odinger equation and the Schr\"odinger algebra. We first
recall the interpretation of non-relativistic holography as
equivalence between representations of the Schr\"odinger algebra
describing  bulk fields  and  boundary fields. One important result
is the explicit construction of the boundary-to-bulk operators in
the framework of representation theory, and  that these operators
and the bulk-to-boundary operators are intertwining operators.
 Further, we recall the fact that there is a hierarchy
of equations on the boundary, invariant w.r.t. Schr\"odinger
algebra. We also review the explicit construction of an
analogous hierarchy  of invariant equations in the bulk, and that
the two hierarchies are equivalent via the bulk-to-boundary
intertwining operators. The derivation of these hierarchies uses a mechanism
introduced first for semi-simple Lie groups and adapted to the non-semisimple
Schr\"odinger algebra. These requires development of the representation theory of the
Schr\"odinger algebra which is reviewed in some detail. We also recall the $q$-deformation
of the Schr\"odinger algebra. Finally, the realization of the Schr\"odinger algebra via
difference operators is reviewed.
\nt {\it PACS numbers:} 11.25.Tq, 02.20.Sv, 03.65.Fd, 03.65.Wj
\end{abstract}

\np

\setcounter{tocdepth}{2}
\tableofcontents

\np

\setcounter{equation}{0}
\section{Introduction}

The role of nonrelativistic symmetries in  theoretical physics was
always important.  Currently one of the  most popular fields in theoretical physics
- string theory, pretending to be a universal theory - encompasses
together relativistic quantum field theory, classical gravity, and
certainly, nonrelativistic quantum mechanics, in such a way that
it is not even necessary to separate these components.

Since the cornerstone of quantum mechanics is the Schr\"o\-din\-ger equation then
 it is not a surprise that the Schr\"o\-din\-ger group -
  the group that is the maximal group of
symmetry of the Schr\"o\-din\-ger equation - is playing recently more and
more a prominent role in theoretical physics. Especially important are the recent developments
to non-relativistic conformal holography,   cf., e.g.,
\cite{NisSon,SakYos,Son,BaMcG,Goldberger,HRR,MaMaTa,ABMcG,DuHaHo,Yamada,HarYos,Schvel,AADV,AADVa,AMSV,BagGop,FueMor,VolWen,ColYava,Bobev,OogPar,AizDob,Naka,RosSar,ImSi,LeiNoa,Davody,HorMel,Brattan,AHOW,Guica,CLZ,KraPer,Berg,Ross,IshNis,Vasil,vanRees,ZZW,KSTa,BTD}.

The latter is natural since originally the Schr\"odinger group, actually the Schr\"odinger
algebra, was introduced by Niederer \cite{Nie} and Hagen \cite{Ha}
as a nonrelativistic limit of the vector-field realization of the
conformal algebra (see also \cite{BX}).

 Recently, Son \cite{Son} proposed another method of identifying the
Schr\"odinger algebra in d+1 space-time. Namely, Son started from
anti de Sitter (AdS) space in d+3 dimensional space-time with metric
that is invariant under the corresponding conformal algebra
so(d+1,2) and then deformed the AdS metric to reduce the symmetry to
the Schr\"odinger algebra.

 In view of the relation of the conformal and Schr\"odinger
algebra there arises the natural question.  Is there a
nonrelativistic analogue of the AdS/CFT correspondence (CFT stands
for Conformal Field Theory), in which the conformal symmetry is
replaced by Schr\"odinger symmetry.  Indeed, this is to be expected
since the Schr\"odinger equation should play a role both in the bulk
and on the boundary.

 Thus, we review some nonrelativistic aspects of the AdS/CFT correspondence.
 Since the literature on the subject is immense we
 mention some aspects that will not be covered in this review:\\
--- Applications of supersymmetry to non-relativistic   holography, cf., e.g.,
\cite{DuvHors,SakYosu,SakYosua,Horvathy,Nakas,DonGau,DonGaua,NSY,LLL,BoKuPi,JKCY,KOSS,HPV,Sour,HPZ,YodNoj,LSZM}.\\
--- Other approaches to non-relativistic   holography, cf., e.g.,
\cite{MSW,LinWu,Taylor,KST,BarFue,Filev,HerYar,KisTim,ORU,Singh,AdaWan,KKT,KYY,LMZ,GPZ,Son:2012,MSD,KWW,JanKar}.\\
--- Applications of non-relativistic holography to condensed matter systems, cf.,  e.g.,
\cite{BSH,RRST,Ammon,BMM,CMST,TonWon,Wang,SWZ,DHIS}.\\
--- Approaches using the Galilean subgroup of the Scr\"odinger  group, cf., e.g.,
\cite{DBPK,DuvHor,SNYY,MarTac,CdBDY,HosRou,HosRoua,GomKam}.\\
--- Applications to the problem of ageing using
various subgroups of the \SG, cf., e.g., \cite{HSSU,HSSUa,MinPle,JLMP,HenSto,MWW,StoHenb,StoHenba}.

Returning to our main topic we first remind the two
ingredients of the AdS/CFT correspondence \cite{Maldacena,GuKlPo,Witten}:\\
1. the holography principle, which is very old, and  means the
reconstruction of some objects in the bulk (that may be classical or
quantum) from some objects on the boundary;\\  2. the reconstruction
of quantum objects, like 2-point functions on the boundary, from
appropriate actions on the bulk.

The realization of the first ingredient is reviewed in Section 3    in  the
simplest case of the (3+1)-dimensional bulk. It is shown that the
holography principle is established using representation theory
only, that is,  no action is specified.
(Such representation-theoretic intertwining operator realization of the AdS/CFT correspondence
in the conformal case was given in \cite{Dobrev}.)

For the implementation of the first ingredient in the Schr\"odinger
algebra context  in \cite{AizDob} was used a method that is standard in the mathematical
literature for the construction of discrete series representations
of real semisimple Lie groups \cite{Hotta,Schmid}, and which method
was applied in the physics literature first in \cite{DMPPT} in
exactly an AdS/CFT setting, though that term was not used then.

The method  utilizes the fact that in the bulk the Casimir operators
are not fixed numerically. Thus, when a vector-field realization of
the algebra in consideration is substituted in the Casimir it turns
into a  differential  operator.  In contrast, the boundary Casimir
operators are fixed by the quantum numbers of the fields under
consideration. Then the bulk/boundary correspondence forces an
eigenvalue equation involving the Casimir differential operator.
That eigenvalue equation is used to find the two-point Green
function in the bulk which is then used to construct the
boundary-to-bulk integral operator. This operator maps a boundary
field   to a bulk field similarly to what was done in the conformal
context by Witten  \cite{Witten} (see also \cite{Dobrev}). This is the  first main result of
\cite{AizDob}.

The second main result of \cite{AizDob} is  that this
operator is an intertwining operator, namely, it intertwines the two
representations of the Schr\"odinger algebra acting in the bulk and
on the boundary. This also helps us to establish that each bulk
field has actually two bulk-to-boundary limits. The two boundary
fields have conjugated conformal weights ~$\Delta$, $3-\Delta$, and
they are related by a  boundary two-point function.

In Section 4 we review the second ingredient of the non-relativistic
version of the AdS/CFT correspondence. Namely, we review the
connection of the results \cite{Son,BaMcG,FueMor} with
 the formalism of \cite{AizDob}.

In Section 5 is reviewed the Schr\"odinger equation as
an invariant differential equation on the (1+1)-dimensional case. On the boundary this was done
 in \cite{DDM} (extending the approach in the
semi-simple group setting \cite{Dob}), constructing actually an
infinite hierarchy of invariant differential equations, the first member being the
free heat/Schr\"odinger equation (see also \cite{ADD}).
In Section 5.4 is reviewed the extension of this construction to the bulk
combining techniques from \cite{AizDob} and \cite{DDM}.

In Section 6 is reviewed the Schr\"odinger equation as
an invariant differential equation in the general (n+1)-dimensional case following
\cite{ADDS,DoSt}. The general situation is very complicated and
requires separate study of the cases: ~$n=2N$ and $n=2N+1$.
In Section 7 is reviewed separately and in more detail the (3+1)-dimensional case,
since it is most important for physical applications.
In Section 8 is reviewed the $q$-deformation of the \S\ in the (1+1)-dimensional case, cf. \cite{DDMq}.
In Section 9 are reviewed the difference analogues of the \S\ in the (n+1)-dimensional case, cf. \cite{DDMd}.

\np

\setcounter{equation}{0}
\section{Preliminaries}

\subsection{Schr\"odinger algebra ~$\hs(n)$}

The Schr\"odinger algebra $\ {\cal  S}(n)\ $, ($n\geq 1$),  in ($n$+1)-dimensional
space-time has ~$(n^2 +3n +6)/2$~ generators with the following
non-trivial commutation relations, cf., e.g., \cite{BarRac}: \eqna{sch}
&&[ J_{ij} , J_{k\ell} ] ~=~ \d_{ik} J_{j\ell} +
\d_{j\ell} J_{ik} - \d_{i\ell} J_{jk} - \d_{jk} J_{i\ell} \\
&&[ J_{ij} , P_k ] ~=~ \d_{ik} P_{j} - \d_{jk} P_{i} \\ &&[
J_{ij} , G_k ] ~=~ \d_{ik} G_{j} - \d_{jk} G_{i}  \\ &&[P_t,
G_i] ~=~ P_i  \\ &&[K, P_i] ~=~ -G_i  \\ &&[D, G_i]
~=~ G_i  \\ &&[D, P_i] ~=~ -P_i  \\ &&[D, P_t] ~=~
-2P_t  \\ &&[D, K] ~=~ 2K \\ &&[P_t, K] ~=~ D  \eena
 where ~$J_{ij} = -J_{ji}$, $i,j = 1,2,\ldots,n$, are the
generators of the rotation subalgebra ~$so(n)$, ~$P_i$, $i =
1,2,\ldots,n$, are the generators of the abelian subalgebra ~$t(n)$~
of space translations, ~$G_i$, $i = 1,2,\ldots,n$, are the
generators of the abelian subalgebra ~$\cg(n)$~ of special Galilei
transformations, ~$P_t$~ is the generator of time translations,
~$D$~ is the generator of dilatations (scale transformations), ~$K$~
is the generator of Galilean conformal transformations.

  Actually, most often we shall work with the central extension of the
Schr\"odinger algebra ~$\hat {\cal S}(n) $,  obtained by adding
the central element $M$ to ${\cal S}(n)$ which enters the
additional commutation relations: \beq\label{comut}
 [P_k, G_{\ell}] = \delta_{k\ell} M\ .\eeq
Note that the centre is one-dimensional.
Of course, \eqref{comut} gives also a central extension
of the Galilei subalgebra $\cg(n)$, however, for
$n=1,2$ this is not the full central extension
~${\hat \cg}(n)$~ of $\cg(n)$, since the centre
is $(n+1)$-dimensional in these cases, cf. e.g., \cite{BarRac}.

The centrally extended \S\ for ~$n=3$~ was introduced in \cite{Nie,Ha} (see also \cite{BX})  by deformation and extension
of the standard vector field realization of the
conformal algebra ~$\cc(3)$~ in ~$3+1$ - dimensional space-time.
The resulting vector field realization for arbitrary $n$ is~:
\eqna{veca}
&&P_{j} ~~=~~ \pd_j\\  &
&G_{j} ~~=~~ t\pd_j + Mx_j \\ &
&P_t ~~=~~ \pd_t \\ &
&D ~~=~~ 2t\pd_t + x_j\pd_j +\D \\ &
&K ~~=~~ t^2\pd_t + tx_j\pd_j + {M\over 2} x_j^2
+ t\D \\ &
&J_{jk} ~~=~~ x_k\pd_j - x_j\pd_k\eena
where ~$\pd_t \equiv \pd/\pd_t\,$, ~$\pd_j \equiv \pd/\pd_{x_j}\,$,
summation over repeated indices is assumed,
 ~$\D$~ is a number called the conformal weight (more about will be said below).
We note that (\ref{veca}{f}) may be extended by the matrices of
a finite--dimensional representation ~$\Sigma_{jk}$~ of ~$so(n)$~
which satisfies (\ref{sch}{a}) as follows:
$$J_{jk} ~~=~~ x_k\pd_j - x_j\pd_k ~+~ \Sigma_{jk} \eqno(\ref{veca}{f'}) $$

Now we list the important subalgebras of the Schr\"odinger algebra $\ {\cal  S}(n)\ $:

 The generators ~$J_{ij},P_i$~ form the
~$((n+1)n/2)$--dimensional Euclidean subalgebra ~$\ce(n)$.

The generators ~$J_{ij},P_i,D$~ form the
~$((n+1)n/2 +1)$--dimensional Euclidean Weyl subalgebra ~$\cw(n)$.

The subalgebras ~$\te(n)$ ann $\tw(d)$~ generated by ~$J_{ij},G_i$
and by ~$J_{ij},G_i,D$, resp., are isomorphic to ~$\ce(n)$,
$\cw(n)$, resp.

The generators ~$J_{ij},P_i,G_i,P_t$~ form the
~$((n+1)(n+2)/2)$--dimensional Galilei subalgebra ~$\cg(n)$. The
generators ~$J_{ij},P_i,G_i,K$~ form another
~$((n+1)(n+2)/2)$--dimensional subalgebra ~$\tg(n)$~ which is
isomorphic to the Galilei subalgebra.

The isomorphic pairs mentioned above are conjugated to each
other in a sense explained below.

For the structure of ~$\sch{n}$~ it is also important to note that
the generators ~$D,K,P_t$~ form an ~$sl(2,\bbr)$~ subalgebra.

Obviously  ~${\cal  S}(n)$~ is not semisimple and  has the following
Levi--Malcev decomposition (for $n\neq 2$): \bea  {\cal  S}(n) ~&=&~
\cn(n) ~\niplus~ \cm(n) \\  \cn(n) ~&=&~ t(n) \oplus g(n)\ , \nn\\
&& t(n) = l.s. \{ P_i\}, ~~ g(n) = l.s. \{ G_i\},\nn\\
\cm(n) ~&=&~ sl(2,\bbr) \oplus so(n), \nn\eea with $\cm(n)$ acting
on $\cn(n)$, where
 the maximal solvable ideal ~$\cn(n)$~ is abelian,
while the semisimple
subalgebra (the Levi factor) is  ~$\cm(n)$.

For ~$n=2$~ the maximal solvable ideal $t(n) \oplus g(n) \oplus so(2)$
is not abelian, while the Levi factor
$sl(2,\bbr)$ is simple. Note, however, that many statements below
will hold for arbitrary $n$, incl. $n=2$, if we extend the definition
of  $\cn(n)$ and $\cm(n)$ to the case $n=2$, (which is
natural since $\cm(2)$ is then reductive, and it is well known
that many semisimple structural and representation-theoretic results
hold for the reductive case).

The commutation relations \eqref{sch}  are graded  if we define:
\eqna{graa}
&&{\rm deg}~D ~~=~~ 0\\ &
&{\rm deg}~G_j ~~=~~ 1\\ &
&{\rm deg}~K ~~=~~ 2\\ &
&{\rm deg}~P_j ~~=~~ -1\\ &
&{\rm deg}~P_t ~~=~~ -2\\ &
&{\rm deg}~M ~~=~~ 0\\ &
&{\rm deg}~J_{jk} ~~=~~ 0  \eena
As expected the corresponding grading operator is ~$D$.

For future reference we record also the
following involutive antiautomorphism of the \S~:
\eqn{coj} \om (P_t) ~=~ K , \quad \om (P_j) ~=~ G_j ,
\quad \om (J_{jk}) ~=~ - J_{jk}\,
\quad \om (D) ~=~ D, \quad \om (M) ~=~ M \ .\eeq
Note that this conjugation is transforming the
isomorphic pairs of subalgebras introduced above,
namely, we have:
\eqn{cojd} \om (\ce) ~=~ \te , \quad
\om (\cw) ~=~ \tw , \quad
\om (\cg) ~=~ \tg \ .\eeq

We end this discussion of the general structure of $\hs(n)$ with the
question of invariant scalar products.
Since the \S\ is not semisimple its Cartan-Killing form
is degenerate. More than that - the \S\ does not
have  any  nondegenerate ad-invariant
symmetric bilinear form \cite{DDM}.   For the discussion
of non-semisimple Lie algebras with nondegenerate ad-invariant
symmetric bilinear form we refer to \cite{MR}.

\bigskip

{\bf Matrix representation}

It is useful to have a representation of $ \hs(n) $ by $(2n+2)\times (2n+2)$
matrices: \bea
  & & D_{ab} = \sum_{\mu=1}^n(-\delta_{a,2\mu}\delta_{b,2\mu} + \delta_{a,2\mu+1} \delta_{b,2\mu+1}),
   \quad  M_{ab} = -2 \delta_{a1} \delta_{b,2n+2}, \nn \\
  & & (G_k)_{ab} = \delta_{a1} \delta_{b,2k}-\delta_{a,2n+3-2k}\delta_{b,2n+2}, \nn \\
  & & (P_k)_{ab} = \delta_{a1} \delta_{b,2n+3-2k} + \delta_{a,2k} \delta_{b,2n+2},
      \label{Repsn} \\
  & & (P_t)_{ab} = -\sum_{\mu=1}^n \delta_{a,2\mu} \delta_{b,2n+3-2\mu},
    \quad (K)_{ab} = \sum_{\mu=1}^n \delta_{a,2\mu+1} \delta_{b,2n+2-2\mu}, \nn \\
  & & (J_{kl})_{ab} = \delta_{a,2l}\delta_{b,2k}-\delta_{a,2k}\delta_{b,2l}
       -\delta_{a,2n+3-2k}\delta_{b,2n+3-2l} + \delta_{a,2n+3-2l}\delta_{b,2n+3-2k}, \nn
\eea where $X_{ab}$ denotes the $(a,b)$ element of matrix $X$. In
this representation $D$ is diagonal : $ D =
diag(0,-1,1,-1,1,\cdots,-1,1,0)$, while $P_t$ and $K$ are
minor-diagonal.

\bigskip
For ~$n=3$~ we give the above  explicitly. Positions of non-zero
entries are indicated by the name of the generators as follows: \beq
n=3, \qquad \left(
\begin{array}{cccccccc}
0 & G_1    & P_3    & G_2    & P_2    & G_3    & P_1    & M   \\
  & D      &        & J_{12} &        & J_{13} & P_t    & P_1 \\
  &        & D      &        & J_{23} & K      & J_{13} & G_3 \\
  & J_{12} &        & D      & P_t    & J_{23} &        & P_2 \\
  &        & J_{23} & K      & D      &        & J_{12} & G_2 \\
  & J_{13} & P_t    & J_{23} &        & D      &        & P_3 \\
  & K      & J_{13} &        & J_{12} &        & D      & G_1 \\
0 &        &        &        &        &        &        & 0
\end{array}
\right)_.                          \label{nonzero3} \eeq The first
column and eighth row are empty.

\subsection{Triangular decomposition of ~$\hs(n)$}

The grading \eqref{graa} can be viewed as extension of the triangular
decomposition of the algebra ~$sl(2,\bbr) ~=~
sl(2,\bbr)^+ \oplus sl(2,\bbr)^0 \oplus sl(2,\bbr)^-$,
~where $sl(2,\bbr)^+$ is spanned by $K$,
the Cartan subalgebra $sl(2,\bbr)^0$ is spanned by $D$,
and $sl(2,\bbr)^-$ is spanned by $P_t\,$. Taking into account
also the triangular decomposition:
~$so(n) ~=~ so(n)^+ \oplus so(n)^0 \oplus so(n)^-$,
(more precisely of its complexification $so(n,\bbc)$),
we can introduce the following triangular decomposition:
\bea\label{trg}  &&\sch{n} ~=~
\sch{n}^+ ~\oplus ~\sch{n}^0 ~\oplus ~\sch{n}^- \\ &&\sch{n}^+ ~=~
g(n) ~\oplus ~sl(2,\bbr)^+
 ~\oplus ~so(n)^+ \nn\\
&&\sch{n}^0 ~=~ sl(2,\bbr)^0
 ~\oplus ~so(n)^0 ~\oplus ~ {\rm lin.span}\ M\ ,\nn\\
&&\sch{n}^- ~=~ t(n) ~\oplus ~sl(2,\bbr)^-
 ~\oplus ~so(n)^- \nn \eea  (Clearly, for $n=1$ the $so(n)$ factors are missing, while for
$n=2$ only the Cartan subalgebra $so(n)^0$ survives.)

\np

\setcounter{equation}{0}
\section{Non-relativistic holography}

\subsection{Choice of bulk and boundary}

In the beginning of this we Section we review Son \cite{Son}.
To realize the Schr\"odinger symmetry in ~$(n+1)$ dimensions geometrically, Son takes the
AdS metric, which is is invariant under the conformal group ~$O(n+2,2)$~ in ~$(n+2)$ dimensions, and
then deforms it to reduce the symmetry down to the Schr\"odinger group.
The AdS space, in Poincar\'e coordinates, is
\eqn{sonds2}
  ds^2 = \frac{\eta_{\mu\nu}dx^\mu dx^\nu + dz^2}{z^2}\ ,~\mu,\nu = 0,1,\ldots,n+1 \ , ~\eta_{\mu\nu} = {\rm diag}(-1,1,\ldots,1)\ .
\eeq
The generators of the conformal group correspond to the following
infinitesimal coordinates transformations that leave the metric
unchanged,
\eqnn{conf-act}
  P^\mu &: &~ x^\mu \to x^\mu + a^\mu, \\
  D &: &~ x^\mu \to (1-a) x^\mu, \quad z \to (1-a)z,\nn\\
  K^\mu &: &~ x^\mu \to x^\mu + a^\mu (z^2 + x \cdot x)
          - 2x^\mu (a\cdot x) \nn
\eea
(here $x\cdot x\equiv\eta_{\mu\nu}x^\mu x^\nu$).

Then Son deforms the above metric so to reduce the symmetry to the
Schr\"odinger group.  The resulting metric is \cite{Son}:
\begin{equation}\label{schroed-met}
  ds^2 =  -\frac{2(dx^+)^2}{z^4} + \frac{-2dx^+ dx^- + dx^i dx^i + dz^2}{z^2}
  \ ,~ i=1,\dots,n \ .
\end{equation}
It is straightforward to verify that the metric~(\ref{schroed-met})
exhibits a full Schr\"odinger symmetry. Indeed, the generators of the
Schr\"odinger algebra correspond to the following isometries of the
metric:
\bea
  P_i :& ~ x_i \to x_i + a_i,\quad
  H: ~ x_+ \to x_+ + a,  \quad M: ~ x_- \to x_- + a, \\
  G_i :& ~   x_i \to x_i - a_i x_+, \quad x_- \to x_- - a_i x_i,\nn\\
  D :& ~ x_i \to (1-a)x_i, \quad z\to (1-a)z, \quad
         x_+ \to (1-a)_2 x_+, \quad x_- \to x_-,\nn\\
  K: & ~ z \to (1 - ax_+) z, \quad x_i \to (1 - ax_+) x_i, \quad
      x_+ \to (1 - ax_+) x_+, \nn\\
     &\qquad   x_- \to x_- - \frac a2 (x_i x_i + z^2)\ ,\nn
     \eea
while the generators $J_{jk}$ of $so(n)$ rotate the coordinates $x_j$ as before.
We require that the Schr\"odinger algebra is an isometry  of the
above metric.  We also need to replace the central element $M$ by
the derivative of the variable $x_-$ which is chosen so that
$\frac{\partial}{\partial x_-}$ continues to be central. Note the variable $x_-$ does not scale
w.r.t. $D$. Such variables are called ultralocal.

Thus, a vector-field realization of the Schr\"odinger algebra ~$\hat {\cal S}(n) $~
 in the ~$(n+3)$-dimensional bulk space ~$(t,x_i, x_-, z)$~ is:
\bea\label{schr-bulk}
&&P_{j} ~~=~~ \pd_j\\  &
&G_{j} ~~=~~ t\pd_j + mx_j\nn \\ &
&P_t ~~=~~ \pd_t \nn\\ &
&D ~~=~~ 2t\pd_t + x_j\pd_j + z \frac{\partial}{\partial z} \nn\\
&&K ~~=~~ t^2\pd_t + tx_j\pd_j + t z \frac{\partial}{\partial z} + \frac{1}{2} (x_j^2 + z^2)M \nn\\
&&J_{jk} ~~=~~ x_k\pd_j - x_j\pd_k \nn \\
   &&   M ~~=~~  \frac{\partial}{\partial x_-}
\nn
 \eea

We would like to treat the   realization \eqref{veca} as
vector-field realization on  the boundary of the  bulk space  ~$(t,x_i, x_-, z)$.
Obviously,  the variable $z$ is the variable distinguishing the
bulk, namely, the boundary is obtained when $z=0$. The exact map
will be displayed below but heuristically, passing from
\eqref{schr-bulk} to \eqref{veca} one first replaces
$z\frac{\partial}{\partial z}$ with $\D$ and then sets $z=0$.

\subsection{One-dimensional case}

Here and below we review the paper \cite{AizDob}.
Now we restrict to the 1+1 dimensional case,
$n=1$. In this case the centrally extended Schr\"o\-din\-ger algebra
has six generators:\\
\bu time translation: $ P_t$ \\
\bu space translation: $ P_x$\\
\bu Galilei boost: $ G$\\
\bu dilatation: $ D$\\
\bu conformal transformation: $ K$\\
\bu mass: $ M$\\
with the following non-vanishing commutation relations:
\beq\label{scha}
  \begin{array}{lclcl}
    [P_t, D] = 2P_t, & & [D, K] = 2K, & & [P_t, K] = D, \\[3pt]
  [P_t, G] = P_x,    & & [P_x, D] = P_x,   & &  \\[3pt]
   [P_x, K] = G,   & & [D, G] = G,   & & \\[3pt]
[P_x, G] = M, & && &
  \end{array}
\eeq

Further we need also the Casimir operator. It turns out that the
lowest order nontrivial Casimir operator is the 4-th order one
 \cite{Perroud}:
 \beq  \tilde{C}_4 = ( 2 M D - \{ P_x, G \} )^2 \ -  \ 2 \{ 2MK - G^2, 2MP_t-P_x^2 \}
  \eeq
 In fact, there are many cancellations, and the central
generator $M$ is a  common linear multiple. (This is seen
immediately by setting $M=0$, then  $\tilde{C}_4\to 0$.)

The metric \eqref{schroed-met} of the four-dimensional bulk space $(t,x, x_-, z) $ now reads:\footnote{This
metric was given first in \cite{IKOP,Orlando}, prior to \cite{Son}, albeit without relation to Schr\"odinger symmetry.}
  \beq
  ds^2 = - \frac{2 (dt)^2}{z^4} +      \frac{ -2 dt dx_- + (dx)^2 + dz^2 }
            { z^2 }  \label{metric}
\eeq

Accordingly, the vector-field realization of the Schr\"odinger algebra is given by:
\bea\label{schr-bulk-one}
    P_t &=& \frac{\partial}{\partial t}, \qquad
      P_x = \frac{\partial}{\partial x}, \qquad
      M = \frac{\partial}{\partial x_-},
   \nn \\[5pt]
   G &=& t \frac{\partial}{\partial x} + x M ,
   \label{VecF1} \\[5pt]
    D &=& x \frac{\partial}{\partial x} + z \frac{\partial}{\partial z}
        + 2 t \frac{\partial}{\partial t},
    \nn \\[5pt]
    K &=& t \left(
         x \frac{\partial}{\partial x} + z \frac{\partial}{\partial z} + t \frac{\partial}{\partial t}
           \right)
        +
        \frac{1}{2} (x^2 + z^2)M   \nn
\eea and it generates an isometry of \eqref{metric}. This
vector-field realization of the Schr\"o\-din\-ger algebra acts on the
bulk fields $\phi(t,x,x_-,z) $.

In this realization the Casimir becomes: \bea
\tilde{C}_4 &=& M^2 C_4, \nn \\[5pt]
  C_4 &=&  \hat{Z}^2 - 4 \hat{Z} - 4 z^2  \hat{S} ~=~ 4 z^2 \del{z}^2 - 8 z \del{z} + 5 - 4 z^2 \hat{S}\ ,
 \label{CasReal2} \\[5pt]
  && \hat{S} \equiv 2 \del{t} \del{x_-}  - \del{x}^2  \ , \label{schr-op}\\[5pt]
  &&\hat{Z} \equiv 2 z \del{z} - 1 \nn
\eea  Note that \eqref{schr-op} is the pro-Schr\"odinger operator.

The  vector-field
realization  \eqref{veca}  of the Schr\"odinger algebra on the boundary becomes:
\bea\label{vecb}
  & & P_t = \frac{\partial}{\partial t}, \qquad
      P_x = \frac{\partial}{\partial x}, \qquad
      M = \frac{\partial}{\partial x_-}\ ,
             \nn \\[5pt]
  & & G = t \frac{\partial}{\partial x} + x M ,
    \\[5pt]
  & & D = x \frac{\partial}{\partial x} +  \D
        + 2 t \frac{\partial}{\partial t},
    \nn \\[5pt]
  & & K = t \left(
         x \frac{\partial}{\partial x} + \D  + t \frac{\partial}{\partial t}
           \right)
        +         \frac{1}{2} x^2 M   \nn
\eea where   $\D$ is the conformal weight.

  Thus, the vector-field
realization of the Schr\"odinger algebra  \eqref{vecb} acts on
the boundary field $\phi(t,x,x_-) $ with fixed conformal weight $\D$.

In this realization the Casimir becomes: \beq \tilde{C}^0_4 = M^2
C^0_4,   \qquad C^0_4 =   (2 \Delta - 1) (2 \Delta - 5)
 \label{CasBou} \eeq
As expected   $ C^0_4 $ is a constant which has the same value if we
replace $\D$ by $3-\D$: \beq C^0_4 (\D) = C^0_4 (3-\D)
\label{pequi}\eeq  This already means that the two boundary fields
with conformal weights $\D$ and $3-\D$ are related, or in
mathematical language, that the corresponding representations are
(partially) equivalent.

\subsection{Boundary-to-bulk  correspondence}

 As we explained
in the Introduction we   concentrate on one aspect of    AdS/CFT
\cite{GuKlPo,Witten}, namely, the holography principle, or
boundary-to-bulk correspondence,  which means to have an operator
which maps a boundary field $\varphi$ to a bulk field
$\phi$,    cf.   \cite{Witten}, also \cite{Dobrev}.

Mathematically, this means the following. We treat both the boundary
fields and the bulk fields as representation spaces of the
Schr\"odinger algebra. The action of the Schr\"odinger algebra in
the boundary, resp. bulk, representation spaces is given by formulae
\eqref{vecb}, resp. by formulae  \eqref{VecF1}. The
boundary-to-bulk operator maps the boundary  representation space to
the   bulk representation space.

The fields on the boundary are fixed by the value of the conformal
weight $\D$, correspondingly, as we saw, the Casimir has the
eigenvalue determined by $\D$: \bea C^0_4 \varphi(t,x,x_-) &=&
\lambda\varphi(t,x,x_-) \ , \\  \lambda &=& (2 \Delta - 1) (2
\Delta - 5) \nn\eea

Thus, the first requirement for the corresponding field on the bulk\\
$\phi(t,x,x_-,z) $ is to satisfy the same eigenvalue equation,
namely, we require:   \bea\label{eigen} C_4 \phi(t,x,x_-,z) &=&
\lambda\phi(t,x,x_-,z) \ , \\ \lambda &=& (2 \Delta - 1) (2 \Delta
- 5) \nn\eea where $C_4$ is the differential operator
given in \eqref{CasReal2}. Thus, in the bulk the eigenvalue
condition is a differential equation.

The other condition is the behaviour of the bulk field when we
approach the boundary: \bea\label{BouBeh}
&   \phi(t,x,x_-,z) \ \rightarrow \ z^\a
   \varphi(t,x,x_-) \ ,\\ &\qquad \a = \D,3-\D
   \nn
 \eea

Let us denote by ~$\hat{C}^\a$~ the space of bulk functions $\phi(t,x,x_-,z) $ satisfying \eqref{eigen} and
\eqref{BouBeh}.

To find the boundary-to-bulk operator we  first find the two-point
Green function in the bulk solving the differential equation: \beq
   (C_4 - \lambda)\, G(\chi, z \,;\, \chi', z')
   =
   z'^4 \, \delta^3(\chi - \chi') \, \delta(z-z')
   \label{Gdef}
\eeq where $ \chi = (t, x_-, x). $

 It is important to use an invariant variable
which here is: \beq\label{udef}
   u =\frac{4 z z'}{ (x-x')^2 - 2(t - t')(x_- - x'_{-} ) +
   (z+z')^2}
  \eeq
The normalization is chosen so that for coinciding points we have $u=1$.

 In terms of $u$ the Casimir becomes: \beq
   C_4  = 4 u^2 (1-u) \frac{d^2}{du^2} - 8 u \frac{d}{du} + 5
   \label{Casimirinu}
\eeq

The eigenvalue  equation can be reduced  to the hypergeometric
equation  by the substitution: \beq  G(\chi, z; \chi', z') ~=~ G(u)
~=~ u^{\alpha} {\hat G}(u)  \eeq and the two solutions are: \beq
 {\hat G}  (u) =  F(\a,\a-1;2(\a-1);u) \ , \quad \a=\D,3-\D \eeq
where $ F =\ _2F_1 $ is the standard hypergeometric function.

As expected at $u=1$ both solutions are singular:
 by \cite{BaEr},  they can be recast into:
  $$
     G(u) = \frac{ u^{\Delta} }{ 1-u } F(\Delta-2,\Delta-1;2(\Delta-1);u),
\quad \a=\D,  $$
    $$
    G(u) = \frac{u^{3-\Delta}}{1-u} F(1-\Delta,2-\Delta;2(2-\Delta);u),
    \quad \a=3-\D \ .
  $$

Now the boundary-to-bulk operator is obtained from the two-point
bulk Green function by bringing one of the points to the boundary,
however, one has to take into account all info from the field on the
boundary.

 More precisely,  we express the function in
the bulk with boundary behaviour \eqref{BouBeh} through the function
on the boundary by the formula: \beq
  \phi(\chi,z) = \int d^3\chi' \, S_\a(\chi-\chi',z) \, \varphi(\chi'),
  \label{Bo2Bu}
\eeq where $ d^3 \chi' = dx'_+ dx'_- dx' $ and  $ S_\a(\chi-\chi',z)
$ is defined  by \beq    S_\a(\chi-\chi',z) = \lim_{z' \rightarrow
0} z'{}^{-\a} \, G(u)
   ~ =~ \left[
        \frac{ 4z }{ (x-x')^2 - 2(t-t') (x_- - x'_-) + z^2 }
     \right]^{\a} \eeq

An important ingredient of this approach is that the bulk-to-boundary
and boundary-to-bulk operators are actually intertwining operators.
To see this we need some more notation.

Let us denote by $L_\a$ the bulk-to-boundary operator : \beq (L_\a \
\phi) (\chi) \doteq \lim_{z \rightarrow 0}z^{-\a}\phi(\chi,z),
\label{Buboug} \eeq where ~$\a = \D,3-\D$ ~consistently with
\eqref{BouBeh}. The intertwining property is: \beq L_\a\circ \hX =
\tX_\a \circ L_\a , \qquad X\in\sch{1}, \label{inta} \eeq where
$\tX_\a$ denotes the action of the generator $X$ on the boundary
\eqref{vecb} (with $\D$ replaced by $\a$ from \eqref{BouBeh}),
$\hX$ denotes the action of the generator $X$ in the bulk
\eqref{VecF1}.

Let us denote by $\tL_\a$ the boundary-to-bulk operator in
\eqref{Bo2Bu}: \beq  \phi(\chi,z) = ( \tL_\a \varphi) (\chi,z)
\doteq
 \int d^3\chi' \, S_\a(\chi-\chi',z) \, \varphi(\chi') \eeq
The intertwining property now is: \beq \tL_\a\circ \tX_{3-\a} = \hX
\circ \tL_\a , \qquad X\in\sch{1}. \label{intb} \eeq

 Next we check consistency of the bulk-to-boundary and
boundary-to-bulk operators, namely, their consecutive application in
both orders should be the identity map: \bea L_{3-\a} \circ \tL_{\a}
&=& {\bf 1}_{\rm boundary}, \label{boubu}\\
\tL_{\a} \circ L_{3-\a} &=& {\bf 1}_{\rm bulk}. \label{bubou}\eea

Checking \eqref{boubu} in \cite{AizDob} was obtained:  \beq
(L_{3-\a} \circ \tL_{\a}\,\varphi) (\chi) = 2^{2\alpha} \pi^{3/2} \,
\frac{ \Gamma(\alpha-\frac{3}{2}) }{ \Gamma(\alpha) }
   \, \varphi (\chi)  \eeq Thus, in order to obtain \eqref{boubu}
 exactly, we have to normalize, e.g., $\tL_{\a}$.

 We note the excluded values ~$\a - 3/2 \notin  \bbz_-$~ for which
 the two intertwining operators are not inverse to each other. This
 means that at least one of the representations is reducible.
This reducibility was established \cite{DDM} for the associated
Verma modules with lowest weight determined by the conformal weight
$\D$ and is reviewed below.

 Checking \eqref{bubou} is now straightforward, but also fails  for
 the excluded values.

Note that checking \eqref{boubu} we used \eqref{Buboug} for $\a \to\
3-\a$, i.e., we used one possible limit of the bulk field
\eqref{Bo2Bu}. But it is important to note that this bulk field has
also the boundary as given in \eqref{Buboug}. Namely, we can
consider the field:
 \beq \varphi_0 (\chi) \doteq (L_\a \ \phi) (\chi) =
\lim_{z \rightarrow 0}z^{-\a}\phi(\chi,z), \label{Bubougg} \eeq where
$\phi(\chi,z)$ is given by \eqref{Bo2Bu}. We obtain immediately:
\beq \varphi_0 (\chi) = \int d^3\chi' \, G_\a(\chi-\chi') \,
\varphi(\chi'),
  \label{Bo2Bu2} \eeq
where \beq G_\a(\chi) = \left[
        \frac {4}{ x^2 - 2t x_- }
     \right]^{\a}.
   \label{Gdef2}\eeq
If we denote by $G_\a$ the operator in \eqref{Bo2Bu2} then we have
the intertwining property: \beq \tX_\a \circ G_\a = G_\a \circ
\tX_{3-\a}\  \ . \label{intc} \eeq Thus, the two boundary fields
corresponding to the two limits of the bulk field are equivalent
(partially equivalent for $\a \in  \bbz + 3/2$). The intertwining
kernel has the properties of the conformal two-point function.

Thus, for generic $\D$ the bulk fields obtained for the two values
of $\a$ are not only equivalent - they coincide, since both have the
two fields $\varphi_0$ and $\varphi$ as boundaries.

\bigskip

\noindent {\bf Remark:} ~For the relativistic AdS/CFT
correspondence the above analysis relating the two fields in
\eqref{Bo2Bu2} was given in \cite{Dobrev}. An alternative treatment
relating these two fields via the Legendre transform was given later
in \cite{KleWit}.

\bigskip

As in the relativistic case there is a range of dimensions when both
fields $\D,3-\D$ are physical: \beq\label{phys} \D_-^0 \equiv 1/2  <
\D < 5/2 \equiv \D_+^0   \ .\eeq  At these bounds
the Casimir eigenvalue $\lambda =
(2\D-1)(2\D-5)$ becomes zero.

\np

\setcounter{equation}{0}
\section{Non-relativistic reduction}

  In this Section we review the connection of \cite{Son,BaMcG,FueMor} with
 the formalism of \cite{AizDob} reviewed in the previous Section.
 For this we consider  the action for a scalar field in the background
(\ref{metric}): \beq
  I(\phi) = -\int d^3\chi dz \sqrt{-g} \,
   ( \partial^{\mu} \phi^* \partial_{\mu}  \phi + m_0^2 |\phi|^2).
  \label{action1}
\eeq
By integrating by parts,  and taking into account a non-trivial
contribution from the boundary, one can see that  $ I(\phi) $ has
the following expression:
\beq
  I(\phi) = \int d^3\chi dz \sqrt{-g} \, \phi^* (\partial^{\mu} \partial_{\mu}  - m_0^2) \phi
  - \lim_{z \rightarrow 0}
    \int d^3\chi \frac{1}{z^3} \phi^* \, z \del{z} \phi.
    \label{action2}
\eeq
The second term is evaluated using (\ref{Bo2Bu}).
For $ z \rightarrow 0, $ one has
\beq
  z \del{z} \phi \ \sim \
  \alpha (4z)^{\alpha} \int d^3\chi'
  \frac{ \varphi(\chi') }{ [(x-x')^2 - 2(x_+-x'_+)(x_- - x'_-)]^{\alpha} }
  + O(z^{\alpha+2}).
  \label{zdphi-bou}
\eeq
It follows that
\bea
 & &
  \lim_{z \rightarrow 0}
    \int d^3\chi \frac{1}{z^3} \phi^* \, z \del{z} \phi
   = \lim_{z \rightarrow 0}  \alpha \int d^3\chi d^3 \chi'
    z^{\a-3} \phi^*(\chi,z) \left( \frac{4}{A} \right)^{\alpha} \varphi(\chi')
 \nn \\[5pt]
 & &
  = 4^{\alpha} \alpha \int d^3\chi d^3 \chi'
    \frac{ \varphi(\chi)^* \varphi(\chi') }{ [(x-x')^2 - 2(x_+ - x'_+)(x_- - x'_-) ]^{\alpha} }.
 \label{action3}
\eea

 The equation of motion being read off from the first term
in (\ref{action2}) can be expressed in terms of the
differential operator (\ref{CasReal2}):
\beq
  (\partial^{\mu} \partial_{\mu}   - m_0^2) \, \phi
  = \left(  \frac{C_4 - 5}{4}  + 2 \partial_{-}^2 - m_0^2  \right) \phi = 0.
  \label{EoM}
\eeq The fields in the bulk (\ref{Bo2Bu}) do not solve the equation
of motion. Now we set an Ansatz for the fields on the boundary: $
\varphi(\chi) = e^{M x_-} \varphi(x_+,x) $. Further we  compactify the $x_-$
coordinate:  $ x_- + a \sim x_- $ as in, e.g., \cite{Goldberger,BarFue}. This leads to a separation of
variables for the fields in the bulk in the following way:
\[
  \phi(\chi,z) =
  e^{Mx_-}
  \int dx'_+ dx' \int_0^a d\xi
  \left(
    \frac{4z}{ (x-x')^2 - 2(x_+-x'_+) \xi + z^2 }
  \right)^{\alpha} e^{-M\xi} \varphi(x'_+,x').
\]
Thus we are allowed to make the identification $ \del{x_-} = M $
both in the bulk and on the boundary \cite{Son,BaMcG}.
We remark that under this identification the operator (\ref{schr-op})
becomes the Schr\"odinger operator.
Integration over $ \xi $ turns out to be incomplete gamma function:
\bea
  & & \phi(\chi,z) = e^{Mx_-} \phi(x_+,x,z),
  \label{nonrel-bulk-fun} \\[7pt]
  & & \phi(x_+,x,z)
      =
      (-2z)^{\alpha} M^{\alpha-1}
      \gamma(1-\alpha, Ma)
  \nn \\[5pt]
  & & \qquad \qquad \quad \times
      \int \frac{ dx'_+ dx' }{ (x_+ - x'_+)^{\alpha} }
      \exp\left( - \frac{ (x-x')^2 + z^2 }{ 2(x_+ - x'_+) }M \right) \,
      \varphi(x'_+,x').
   \label{nonrel-bulk-fun2}
\eea This formula was obtained first in \cite{FueMor}. The equation
of motion (\ref{EoM}) now reads \beq
 \left(
   \frac{\lambda - 5}{4} - m^2
 \right) \phi(x_+,x,z) = 0,
\eeq where  $ m^2 = m_0^2 - 2M^2. $  Requiring $ \phi(x_+,x,z) $ to
be a solution to the equation of motion makes the connection between
the conformal weight and mass: \beq
  \Delta_\pm = \frac{1}{2} ( 3 \pm \sqrt{9 + 4m^2} ).
  \label{Delta2mass}
\eeq This result is identical to the relativistic AdS/CFT
correspondence \cite{GuKlPo,Witten}.  The action (\ref{action2})
evaluated for this classical solutions has the following form
($\a=\D_\pm$): \bea
 & &
  I(\phi) =
    -(-2)^{\alpha} M^{\alpha-1} \a \gamma(1-\alpha,Ma)
 \nn \\[5pt]
 & & \qquad \qquad \times
  \int \frac{dx dx_+ dx' dx'_+}{ (x_+ - x'_+)^{\alpha} }
  \exp\left( - \frac{(x-x')^2}{2(x_+ - x'_+)} M \right) \,
  \varphi(x_+,x)^* \varphi(x'_+,x').
  \label{bou-ansatz}
\eea The two-point function of the operator dual to $ \phi $
computed from (\ref{bou-ansatz})  coincides with the result of
\cite{Son,BaMcG,Henkel,StoHen,StoHena}.
 We remark that the Ansatz for the boundary fields
$ \varphi(\chi) = \exp(Mx_- -\omega x_+ + ikx) $ used in \cite{Son,BaMcG}
is not necessary to derive (\ref{bou-ansatz}).

  One can also recover the solutions in \cite{Son,BaMcG}
rather simply  in the group theoretical context of \cite{AizDob}. We
use again the eigenvalue problem of the differential operator
(\ref{CasReal2}): \beq
   C_4 \, \phi(x_+,x,z) = \lambda \, \phi(x_+,x,z). \label{EVprob}
\eeq
but make separation of variables
\medskip
$ \phi(x_+,x,z) = \psi(x_+,x) f(z). $ Then (\ref{EVprob}) is written as follows:
\[
  \frac{1}{f(z)}
  \left(
     \del{z}^2 - \frac{2}{z} \del{z} + \frac{5-\lambda}{4z^2}
  \right)
  f(z)
  =
  \frac{1}{\psi(x_+,x)} \hat{S} \psi(x_+,x) = p^2 \ \mbox{(const)}
\]
Schr\"odinger part is easily solved:
$ \psi(x_+,x) = \exp(-\omega x_+ + i k x) $ which gives
\beq
  p^2 = -2M \omega + k^2. \label{enegy}
\eeq
The equation for $ f(z) $ now becomes
\beq
  \del{z}^2 f(z) - \frac{2}{z}\, \del{z} f(z) +
  \left(
    2M\omega - k^2 - \frac{m^2}{z^2}
  \right) f(z) = 0.
  \label{eqforz}
\eeq This is the equation given in \cite{Son,BaMcG} for $ d=1$. Thus
solutions to equation (\ref{eqforz}) are given by modified Bessel
functions: $
  f_{\pm}(z) = z^{3/2} K_{\pm \nu}(pz)
$ where $ \nu $ is related to the effective mass $m$
\cite{Son,BaMcG}.  In the group theoretic approach one can see its
relation to the eigenvalue of $C_4: $ $ \nu = \sqrt{\lambda+4}/2$
\cite{AizDob}.

  We close this section by giving the expression of
(\ref{bou-ansatz}) for the alternate boundary field
$ \varphi_0. $ To this end, we again use the Ansatz
$ \varphi(\chi) = e^{M x_-} \varphi(x_+,x) $ for (\ref{Bo2Bu2}).
Then performing the integration over $ x'_- $
it is immediate to see that:
\beq
 \varphi_0(x,x_+) \sim
 e^{Mx_-} \int \frac{dx' dx'_+}{ (x_+ - x'_+)^{\alpha} }
  \exp\left( - \frac{(x-x')^2}{2(x_+ - x'_+)} M \right) \,
  \varphi(x'_+,x').
  \label{Bo2Bo-NR}
\eeq
One can invert this relation since
$ G_{3-\alpha} \circ G_{\alpha} = 1_{\rm boundary}. $
Substitution of (\ref{Bo2Bo-NR}) and its inverse to (\ref{bou-ansatz})
gives the following expression:
\beq
  I(\phi) \sim
  \int \frac{dx dx_+ dx' dx'_+}{ (x_+ - x'_+)^{3-\alpha} }
  \exp\left( - \frac{(x-x')^2}{2(x_+ - x'_+)} M \right) \,
  \varphi_0(x_+,x)^* \varphi_0(x'_+,x').
  \label{bou-ansatz2}
\eeq

\np

\setcounter{equation}{0}
\section{Non-relativistic invariant differential equations for ~$n=1$}

\subsection{Canonical procedure}
  In this subsection, we briefly outline the method of \cite{Dob} that shall be used
subsequently. Let $G$ be a complex semisimple Lie group
and ${\mathfrak g}$ its Lie algebra. Let  $ {\mathfrak g} =
{\mathfrak g}_+ \oplus {\mathfrak g}_0 \oplus {\mathfrak g}_- $
be the standard triangular decomposition. We consider
representations of $ {\mathfrak g} $ whose representation spaces
${\cal C}_{\Lambda}$ are $ {\mathbb C}^{\infty} $ functions $ {\cal F} $ on $
G$ with the property called $right\ covariance$
\begin{equation}
   {\cal F}(g x g_-) = e^{\Lambda(H)} {\cal F}(g), \quad g\in G,\
   x = e^H \in G_0, \ H \in {\mathfrak g}_0,\ g_-\in G_- \ , \label{rc}
\end{equation}
where $\Lambda\in {\mathfrak g}_0^*\,$,
$ \Lambda(H) \in {\mathbb Z}$, $G_0 \equiv \exp {\mathfrak
g}_0\,$, $G_\pm\equiv \exp {\mathfrak g}_\pm\,$. Thus the
functions of ${\cal C}_{\Lambda}$ are actually functions on the coset
$G/B$, $B\equiv G_0 G_-$ being a Borel subgroup, or
equivalently\cite{Dob}, on $G_+$ which is dense in $G/B$. The
restricted representation spaces of functions $f$ on $G_+$, such that
$ f = {\cal F}|_{G_+} $, we denote by $ C_{\Lambda} $.
We introduce the left $ \pi_L(X) $ and right $ \pi_R(X) $
actions of $\mathfrak g$ on $ {\cal C}_\Lambda $ by the standard formulae
\begin{equation}
 \pi_L(X) {\cal F}(g) \equiv \left. \frac{d}{d\tau} {\cal
F}(e^{-\tau X} g) \right|_{\tau=0}\ \ ,  \quad
 \pi_R(X) {\cal F}(g) \equiv \left. \frac{d}{d\tau} {\cal F}(g
e^{\tau X}) \right|_{\tau=0}\ \ ,
 \label{LRaction}
\end{equation}
where  $ X \in \mathfrak g\,$. The left action gives
representations of $ \mathfrak g $ by differential operators. It
may be considered for arbitrary $\Lambda(H)\in {\mathbb C}$
and it may be restricted to $ C_\Lambda $. On
the other hand, the space $ {\cal C}_{\Lambda} $ has a lowest
weight structure with respect to the right action, since one can show,
using the right covariance, that
\begin{equation}
  \pi_R(H) {\cal F}(g) = \Lambda(H) {\cal F}(g), \qquad \pi_R(X)
{\cal F}(g) = 0, \label{LWstr}
\end{equation}
where $H \in {\mathfrak g}_0 $ and $ X \in {\mathfrak g}_-\,$.
Thus we are prompted to employ properties of the Verma module $
V^{\Lambda} $ with the lowest weight $ \Lambda\,$,
such that $ V^{\Lambda} \simeq U({\mathfrak g}_+) v_0\,$,
 where $v_0$ is the lowest weight vector,
$ U({\mathfrak g}_+) $
is the universal enveloping algebra of $ {\mathfrak g}_+\,$.
When a Verma module is reducible, it has (at least one) singular
vector $v_s$ such that:\ $H v_s = \Lambda' (H)$, $\Lambda'\neq
\Lambda$, $H \in {\mathfrak g}_0 $, $X v_s =0$, $ X \in
{\mathfrak g}_-\,$. It has the general structure  $ v_s = P({\mathfrak g}_+)
v_0\,$,  where $ P({\mathfrak g}_+) $ is a
polynomial of the generators of $ {\mathfrak g}_+\,$. Then it is
shown that the same polynomial $ P({\mathfrak g}_+) $ gives rise
to a ${\mathfrak g}$-invariant differential equation given explicitly by:
$ P(\pi_R({\mathfrak g}_+)) \psi = 0 $, $\psi = {\cal F},\ f\,$.

Below the procedure of \cite{Dob} shall be used in our non-semisimple Schr\"odinger setting.

\subsection{Verma modules and singular vectors}

In this subsection we follow \cite{DDM}, see also \cite{ADD,ADDS,DoSt}. We consider  lowest
weight modules (LWM) over $\sch{n}$, in particular, Verma modules,
which are standard for semisimple Lie algebras (SSLA) and their
$q$--deformations. For more information on representations of the \S\ we refer to
\cite{RidWin,HenUnt,FeKoSc,FeKoSca,Berce,Camp,Unter,Stoimenov,BagMan,Aiz,Aiza,Aizb,AizIsa,Dong,WuZhu,WuZhua,AGM,AGMa,AnGo,FeiSch,FeiScho}.

A lowest weight module (LWM) $M^\L$ over ~$\sch{n}$~
is given by the lowest weight $\L\in\ch^*$ ($\ch^*$ is the dual of
$\sch{n}^0$) and a lowest weight vector $v_0$ so that $Xv_0 = 0$ if $X\in\sch{n}^-$,
~$Hv_0 = \L (H) v_0$~ if ~$H\in \sch{n}^0$. In particular, we use the Verma modules
$V^\L$ over ~$\sch{n}$~ which are the
lowest weight modules induced from a one--dimensional representation of  the analogue of a Borel subalgebra
~$\cb ~\def~ \sch{n}^0 \oplus \sch{n}^-$
spanned by $v_0$. The Verma module is given explicitly by ~$V^\L ~\cong~
U(\sch{n}^+)~\otimes ~v_0\,$, where ~$U(\sch{n}^+)$~ is  the  universal
enveloping algebra of $\sch{n}^+$.
Further, for brevity we shall omit the sign ~$\otimes$~, ~i.e., we
shall write instead of ~$\otimes v_0$~ just ~$v_0\,$.

Now we restrict to the case ~$n=1$. Then the Cartan subalgebra ~$\sch{1}^0$~ is generated by ~$D,M$~
and we can write all
above mentioned properties as: \bea\label{bact} && D~v_0 ~=~ \D~v_0 ~, \quad
M~v_0 ~=~ M~v_0 ~, \\
&& P_x~v_0 ~~=~~ 0, \quad P_t~v_0 ~~=~~ 0 \nn\eea where ~$\D\in\bbr$~
is the (conformal) weight.

The Borel subalgebra  ~$\cb$~ is generated by the nonpositively graded generators
 ~$D,M,P_x,P_t$. Now we denote the
Verma module as ~$V^\D$~ since ~$M$~ is constant.

   Clearly, ~$U(\sc^+)$~ is abelian and
has basis elements ~$p_{k,\ell} ~=~ G^k K^\ell$. The basis vectors
of the Verma module are ~$v_{k,\ell} ~=~ p_{k,\ell} \otimes v_0$,
(with $v_{0,0} = v_0$). The action of  $\sc$\ on this basis is
derived easily from \eqref{scha}: \bea\label{vact} &&D~v_{k,\ell}
~~=~~ (k + 2\ell +\D)~ v_{k,\ell}  \cr &&G~v_{k,\ell} ~~=~~
v_{k+1,\ell}
 \cr &&K~v_{k,\ell} ~~=~~ v_{k,\ell+1}  \cr
&&P_x~v_{k,\ell} ~~=~~ \ell ~v_{k+1,\ell-1} ~+~ M\, k ~v_{k-1,\ell}
 \cr &&P_t~v_{k,\ell} ~~=~~ \ell (k+\ell-1+\D) ~v_{k,\ell-1}
~+~ M\,{k(k-1) \over 2} ~v_{k-2,\ell}  \eea

Because of \eqref{vact} we notice that the Verma module $V^\D$ can
be decomposed in homogeneous (w.r.t. $\ D$) subspaces as follows:
\bea\label{vdec} &&V^\D ~~=~~ \oplus_{n=0}^\infty ~ V^\D_n \cr
&&V^\D_n ~~=~~ {\rm lin.span.}~\{ v_{k,\ell} ~|~ k+2\ell =n \}
 \cr &&{\rm dim}~ V^\D_n ~~=~~ 1 + \left[{n\over 2}\right]
\eea

Next we analyze the reducibility of ~$V^\D$~ through the so-called
singular vectors. In analogy to the SSLA situation (cf., e.g., \cite{Dob}) a singular vector
~$v_s$~ here is a homogeneous element of ~$V^\D$~, such that
~$v_s\notin \bbc v_0$, and \beq\label{sing} P_x~v_s ~~=~~ 0~,
\quad P_t~v_s ~~=~~ 0 \eeq

All possible singular vectors were given explicitly  in
\cite{DDM}, where was proved:

\nt {\bf Proposition 1.} ~ The singular vectors of the Verma module
~$V^\D$~ over  $\sc$\ are given as follows: \bea\label{sngz} v^p_s
~=&&~ a_0 \sum_{\ell =0}^{p/2} ~ (-2M)^\ell ~{p/2 \choose \ell}~
v_{p-2\ell,\ell} ~= \\ =&&~ a_0 \Bigl( G^2 ~-~ 2MK \Bigr)^{p/2}
\otimes v_0 ~, \quad \D ~=~ {3-p \over 2} ~, \quad p\in 2\bbn  ,
~Ma_0 \neq 0 \cr &&\cr v^p_s ~=&&~ a_0 v_{p0} ~=~ a_0 G^p \otimes
v_0 ~, \quad \D ~{\rm arbitrary}, ~p\in\bbn, ~M=0, ~ a_0 \neq 0 \ . \qquad \dia \nn\eea

\bigskip

\nt
{\bf Remark:}~ We  stress the very different character
of the representations for ~$M\neq 0$~ and ~$M=0$~
from one another and furthermore from the semisimple case.
For ~$M\neq 0$~ and fixed lowest weight at most
one singular vector may exist  and that vector can
be only of ~{\it even} grade.
For  ~$M=0$~ an infinite number of
singular vectors exist - one for each positive grade - and there
is no restriction on the weight. This difference is because
the value $M=0$ changes the algebra - it is not a centrally extended
one anymore. Both cases ~{\it differ from the semisimple case}.
To compare we take, e.g., the algebra $sl(2)$ since it also has only
one Cartan generator as ~$\sch{1}$. For $sl(2)$
for a fixed lowest weight  only one singular vector is possible,
however, for any `grade' $n\beta$, where $n\in\bbn$,
$\beta$ the positive root of $sl(2)$, and not just for even $n$.
~$\diamondsuit$

\bigskip

Whenever there is a singular vector the Verma module
is reducible. We could  analyze this reducibility
also via an analogue of the Shapovalov form \cite{Sha}  used in the
semisimple case. This is a bilinear form  which we define using the
involutive antiautomorphism of the \S\ (cf. \eqref{coj})~:
\eqn{conj} \om (P_t) ~=~ K , \quad \om (P_x) ~=~ G ,
\quad \om (D) ~=~ D, \quad \om (M) ~=~ M \eeq
Explicitly, the form here is given by:
\eqnn{frm}
\left( v_{k\ell}~, ~v_{k'\ell'} \right) ~&=&~
\left( p_{k\ell} \otimes v_0~ , ~p_{k'\ell'} \otimes v_0 \right)
~\equiv~
\left(
v_0 ~, ~\om (p_{k\ell}) ~p_{k'\ell'} ~\otimes ~v_0 \right) ~=\nn\\
&=&~ \left( v_0 ~, ~P_t^\ell ~P_x^k ~G^{k'} ~K^{\ell'} ~v_0 \right)
\eea
supplemented by the normalization condition ~$(v_0, v_0) ~=~ 1$.
Clearly, subspaces with different weights are orthogonal w.r.t.
to this form:
\eqn{ort} \left( v_{k\ell} ~, ~v_{k'\ell'} \right) ~\sim~
\d_{k+2\ell,k'+2\ell'} \eeq
To show this for ~$k+2\ell>k'+2\ell'$~
we move all ~$P_t$ and $P_x$
operators to the right until there are no $G$ and $K$ operators left,
while for ~$k+2\ell<k'+2\ell'$~ we first rewrite the LHS of \eqref{ort} as:
\eqnn{orta}
\left( v_{k\ell} ~, ~v_{k'\ell'} \right) ~=~
\left( p_{k\ell} \otimes v_0 ~, ~p_{k'\ell'} \otimes v_0 \right)
~=&~
\left( \om (p_{k'\ell'}) ~p_{k\ell} ~\otimes ~
v_0 ~, ~v_0 \right) ~=\cr
=&~ \left( P_t^{\ell'} ~P_x^{k'} ~G^{k} ~K^{\ell} ~v_0 ~, ~v_0 \right)
\eea
and then again move all ~$P_t$ and $P_x$
operators to the right until there are no $G$ and $K$ operators left.
The above also shows that the form is symmetric.
In the case ~$k+2\ell ~=~ k'+2\ell'$~ we have the following
explicit expression:
\eqnn{ortb}
\left( v_{k,\ell} ~, ~v_{k+2a,\ell-a} \right) ~=&~
{m^{k+a} \, k! \, \ell! \,(k-d+a)_{\ell-a} \, (k+1)_{2a}
\over 2^a \, a!}
 ~\times \cr &\times ~
_3F_2 ({k\over 2}, {1-k\over 2}, a-\ell ;
1+a, 1-k+d-\ell;1) \eea
where ~$a\in\bbz_+\,$,
~$(a)_p ~\doteq ~\G(a+p)/\G(a)$~ is the Pochhammer symbol and
~$_3F_2(a,b,c;a',b';y)$~ is a generalized hypergeometric series:
\eqn{hyp}_3F_2(a,b,c;a',b';y) ~\doteq~
\sum_{s\in\bbz_+} ~{(a)_s (b)_s (b)_s
\over s! (a')_s (b')_s} ~y^s \eeq
which for ~$a$, or ~$b$, or ~$c$ $\in \bbz_-$~ reduces to a polynomial.

A singular vector is orthogonal to any other vector
w.r.t. to the form \eqref{frm}. Thus we expect (as in the SSLA case)
to obtain the same reducibility results analyzing
the so-called determinant formula.
The determinant formula is the determinant of the matrix ~$\cm_p$ ~of all
Shapovalov forms  at a fixed grade ~$p$. In \cite{DDM} was made the following
~{\it conjecture}~ for ~det~$\cm_p$~:
\eqn{dett}
{\rm det}~\cm_p ~=~ const.(p)~m^{\a_p} ~\prod_{i=0}^{
\left[\textstyle{p\over2}\right]-1}
~(2\D-1+2i)^{\left[\textstyle{p\over2}\right]-i}\ , \eeq
$$\a_p ~=~ \begin{cases} {p^2+2p \over 4} &~for ~$p$ ~even \cr
                  {(p+1)^2 \over 4} &~for ~$p$ ~odd \end{cases}
$$
which there was  verified for ~$p\leq 6$.
We see that the determinant has zeroes exactly in
the cases when we have singular vectors. The above conjecture was proved in \cite{ASK}.

Further we consider the consequences of the reducibility of
the Verma modules. We start with ~$M\neq 0$~ and
consider the subspace of ~$V^{(3-p)/2}$~:
\eqn{sbs} I^{(3-p)/2} ~~=~~ U(\cs^+) ~v^p_s \eeq
It is invariant under the action of the \S. Indeed,
all vectors have grade ~$\geq \deg_{\min} =p$~: ~
lower grades can not be achieved
since the negative grade generators
annihilate ~$v^p_s$. Furthermore this subspace is
isomorphic to a Verma module ~$V^{d'}$ with shifted weight
~$\D' = \D+p = (p+3)/2$. The latter Verma module has no singular
vectors, since its weight is restricted from below~:~ $\D'\geq 5/2$,
while by \eqref{sngz}\ the necessary weight is $\leq 1/2$.

Let us denote the factor--module ~$V^{(3-p)/2} /I^{(3-p)/2}$~
by ~$\cl^{(3-p)/2}$. Let us denote by ~$\vr p\rg$~ the lowest weight
vector of $\cl^{(3-p)/2}$. It satisfies the following conditions:
\eqna{fac}
&&P_x ~\vr p\rg ~~=~~ 0 \\ &
&P_t ~\vr p\rg ~~=~~ 0 \\ &
&\Bigl( G^2 ~-~ 2M K \Bigr)^{p/2} ~\vr p\rg ~~=~~ 0  \eena

Consider in more detail the simplest example ~$p=2$.
In this case the last condition \eqreff{fac}{c} is:
\eqn{faca} G^2 ~\vr 2\rg ~~=~~ 2M K  ~\vr 2\rg \eeq
i.e., ~$K$~ can be replaced by ~$G^2/2M$.
Thus, all vectors of a fixed grade are proportional:
\eqn{dege} G^k ~K^\ell ~\vr 2\rg
~~=~~ {1\over (2M)^\ell }~ G^{k+2\ell}
~\vr 2\rg ~, \quad \D ~=~ 1/2 \eeq
so all graded subspaces are one-dimensional,
i.e., we have a ~{\it singleton}~ basis:
\eqn{dega} {\rm dim}~V^{1/2}_n ~~=~~ 1 ~, \quad \forall n \eeq
which is given only in terms of ~$G$.
(The term `singleton' was used  first for two special
representations of the algebra $so(3,2)$ \cite{Dirac,Fro,Froa,Frob,FlFr,Froc}.)

Analogously, for arbitrary ~$p\in 2\bbn$~ and ~$\D =(3-p)/2$,
from \eqreff{fac}{c} we see that:
\eqn{expr} K^{p/2}  ~\vr p\rg ~~=~~ -
\sum_{\ell=0}^{p/2-1} ~{1\over (-2M)^{p/2-\ell} }
~{p/2 \choose \ell}~
G^{p-2\ell} ~ K^\ell ~\vr p\rg \eeq
Applying repeatedly this relation to the basis one can get rid of all
powers of $K$ which are ~$\geq p/2$. Thus the basis of ~$\cl^{(3-p)/2}$~
will be quasi--singleton if $p\geq 4$, namely,
\eqn{degb} {\rm dim}~V^{(3-p)/2}_n ~~=~~ 1 ~, \quad
{\rm for}~ n=0,1 ~{\rm or} ~n\geq p \eeq
and it is given by:
\eqn{basi} v^p_{k\ell} ~ \equiv ~
G^k ~K^\ell ~\vr p\rg ~, \qquad p\in 2\bbn, ~
k,\ell \in\bbz_+\,, ~\ell \leq p/2 -1, ~\D = {3-p\over 2} \eeq
The transformation of this basis is easily obtained from
\eqref{vact}~:
\eqna{vacta} &
&D~v^p_{k,\ell} ~~=~~ \left(k + 2\ell + {3-p\over 2}\right)~
v^p_{k,\ell} \\
&&G~v^p_{k,\ell} ~~=~~ v^p_{k+1,\ell} \\
&&K~v^p_{k,\ell} ~~=~~
\begin{cases} v^p_{k,\ell+1} &~ \ell < {p\over 2} -1 \cr
- \sum_{s=0}^{p/2-1}  ~{1\over (-2M)^{p/2 -s} }
~{p/2 \choose s} ~v^p_{k+p-2s,s}
&~ \ell = {p\over 2} -1 \cr \end{cases}
\\
&&P_x~v^p_{k,\ell} ~~=~~ \ell ~v^p_{k+1,\ell-1} ~+~
mk ~v^p_{k-1,\ell} \\
&&P_t~v^p_{k,\ell} ~~=~~ \ell \left(k+\ell + {1-p\over 2}\right)
~v^p_{k,\ell-1} ~+~
m{k(k-1) \over 2} ~v^p_{k-2,\ell}  \eena
{}From the transformation rules we see that
~$\cl^{(3-p)/2}$~ is irreducible. It is also clear that
in the simplest case ~$p=2$~ the irrep ~$\cl^{1/2}$~ is
also an irrep of the centrally extended Galilean subalgebra ~$\hhg(1)$~
spanned by ~$P_x,P_t,G$.

For ~$M=0$~ we
consider the subspaces of ~$V^{\D}$~:
\eqn{sbsz} I^{\D}_p ~~=~~ U(\cs^+) ~G^p ~\otimes ~v_0 ~, \qquad p\in\bbn \eeq
They are invariant under the action of the \S, which is shown
as in the case $M\neq 0$. The corresponding singular vectors are:
\eqn{sngc}  \tilde v^p_s ~~=~~ G^p ~\otimes ~v_0 \eeq
Furthermore the subspace ~$I^{\D}_p$~ is
isomorphic to a Verma module ~$V^{\D'}$ with shifted weight
~$\D' = \D+p$. The latter Verma module again has an infinite
number of singular vectors, and an infinite number of
subspaces ~$I^{\D+p}_{p'} ~=~ U(\cs^+) G^{p'} \otimes v_0$,
~$p'\in\bbn$, isomorphic to Verma modules $V^{\D+p+p'}$.
Furthermore, the original Verma module
~$V^{\D}$~ is itself a submodule of an
infinite number of Verma modules $V^{\D+p''}$, $p''\in\bbn$.
Altogether, for each ~$\D$~ there exists a doubly infinite sequence
of Verma modules:~
\eqn{nest} \cdots ~\supset ~V^{\D-1} ~ \supset ~V^{\D} ~ \supset ~
V^{\D+1} ~ \supset ~ \cdots ~, \qquad M=0, ~\D ~{\rm arbitrary} \eeq
Of course, all ~$V^\D$ whose weights differ by an integer are in
one and the same sequence. Such embedding diagrams were called
multiplets in \cite{Domu}.

For each ~$V^\D$~ the submodule ~$I^\D_1 ~\cong
V^{\D+1}$ contains as submodules all other submodules of
~$V^\D$, i.e., in \eqref{sngc} only the singular vector with $p=1$
is relevant.  Consider the factor space ~$\tilde \cl^\D_0 ~=~
V^\D/V^{\D+1}$~ and  denote by
~$\widetilde{\vr 0\rg}$~ its lowest weight vector.
The latter satisfies the following conditions:
\eqna{faco}  &
&P_x ~\widetilde{\vr 0\rg} ~~=~~ 0 \\ &
&P_t ~\widetilde{\vr 0\rg} ~~=~~ 0 \\ &
&G ~\widetilde{\vr 0\rg} ~~=~~ 0 \eena
Consequently, the basis of  ~$\tilde \cl^d_0$~ is given by:
\eqn{baso} \hv^0_\ell ~~=~~ K^\ell ~\widetilde{\vr 0\rg} \eeq
This is another example of an even smaller than singleton basis,
since the odd--graded levels are empty. Thus, we call the basis in
\eqref{baso}\ a ~{\it singleton--void}~ basis.
Its transformation rules  are (cf. \eqref{vact})~:
\eqna{vactb}   &
&D~\hv^0_\ell ~~=~~ (2\ell +\D)~ \hv^0_\ell \\ &
&G~\hv^0_\ell ~~=~~ 0 \\ &
&K~\hv^0_\ell ~~=~~ \hv^0_{\ell+1}]\\ &
&P_x~\hv^0_\ell ~~=~~ 0\\ &
&P_t~\hv^0_\ell ~~=~~ \ell (\ell-1+\D) ~\hv^0_{\ell-1} \eena
Clearly, ~$\tilde \cl^\D_0$~ is in fact a Verma module over the
~$sl(2,\bbr)$~ subalgebra spanned by ~$D,K,P_t$. It is well
known that such a Verma module is reducible (cf., e.g., \cite{Dob})
iff ~$\D\in \bbz_-$. In this case there exists a singular
vector given by:
\eqn{sngs} v^0_s ~~=~~ \hv^0_{1-\D} ~~=~~ K^{1-\D} ~\widetilde{\vr 0\rg}
\,, \quad \D\in \bbz_- \eeq
Thus the invariant subspace ~$\tilde I^\D_0$~ of $\tilde \cl^\D_0$
is spanned by ~$K^{1-\D+\ell} ~\widetilde{\vr 0\rg}$, ~$\ell\in\bbz_+\,$,
and is isomorphic to another Verma module $\tilde \cl^{\D-2}_0$
which is  irreducible.

Let us stress that ~$v^0_s$~ is not a singular vector of the original
Verma module ~$V^\D$, but  of its factor module $\tilde \cl^\D_0$.
 Vectors, which become singular vectors only in
factor modules, are called ~{\it subsingular}~ vectors.

Consider now
the factor space ~$\cl^\D_0 ~=~ \tilde \cl^{\D}_0 /\tilde \cl^{\D-2}_0$,
denoting by ~$\vr 0\rg$~ its lowest weight vector.
It satisfies the following conditions:
\eqna{facp} &
&P_x ~\vr 0\rg ~~=~~ 0\\  &
&P_t ~\vr 0\rg ~~=~~ 0\\  &
&G ~\vr 0\rg ~~=~~ 0\\  &
&K^{1-\D} ~\vr 0\rg ~~=~~ 0\eena
Consequently, the basis of  ~$\cl^\D_0$~ is given by:
\eqn{basp} v^0_\ell ~~=~~ K^\ell ~\vr 0\rg  \,, \quad -\D,\ell \in\bbz_+\,,
 ~\ell\leq -\D \eeq
Hence ~$\cl^\D_0$~ is finite--dimensional: ~dim~$\cl^d_0 = 1-\D$, and
in fact, when ~$\D$~ runs through ~$\bbz_-$~ one obtains
all irreducible finite--dimensional  representations of ~$sl(2,\bbr)$.
The latter are not unitary, except in the trivial one-dimensional case
obtained for $\D=0$. The transformation rules for ~$v^0_\ell$~
are as for ~$\hv^0_\ell$~ in \eqref{vactb}, except that ~$K v^0_\D ~=~ 0$.

Summarizing the above  in \cite{DDM} was proved:

\nt
{\bf Theorem 1.} ~
The list of the irreducible lowest weight modules over the
(centrally extended) \S\ is given by:
{\bu} ~$V^d$~, ~when ~$\D\neq (3-p)/2$, ~$p\in 2\bbn$ and ~$M\neq 0$; \\
{\bu} ~$\cl^{(3-p)/2}$, ~when ~$\D= (3-p)/2$,
~$p\in 2\bbn$ and ~$M\neq 0$;\\
{\bu} ~$\tilde \cl^\D_0$, ~when ~$\D\notin\bbz_-$ and ~$M=0$;\\
{\bu} ~$ \cl^\D_0$, ~when ~$d\in\bbz_-$ and ~$M=0$.

\nt
In the last case one has~:~ dim~$\cl^\D_0 = 1-\D$; in all other
cases the irreps are infinite--dimensional. The representation ~$\cl^{1/2}$~
is also an irrep of the centrally extended Galilean subalgebra ~$\hhg(1)$.
The irreps in the last two cases are also irreps of the
subalgebra ~$sl(2,\bbr)$.

\subsection{Generalized Schr\"odinger equations from a
vector--field realization of the \S}

Now we shall employ vector--field representation \eqref{vecb} as
in \cite{DDM}. This realization was  used to construct a
polynomial realization of the irreducible lowest weight modules
considered in the previous Subsection.  For this realization we
represent the lowest weight vector by the function 1. Indeed, the
constants in \eqref{vecb}  are chosen so that \eqref{bact}\ is
satisfied:
\beq\label{bacf} D~1 ~~=~~ \D ~, \quad M~1 ~~=~~ M ~,
\quad P_x~1 ~~=~~ 0, \quad P_t~1 ~~=~~ 0 \eeq Applying the basis
elements ~ $p_{k,\ell} = G^k K^\ell$~ of the universal enveloping
algebra $U(\sc^+)$ to 1 we get polynomials in ~$x,t$. Let us
introduce notation for these polynomials by ~$f_{k,\ell} ~\equiv ~
p_{k,\ell}~1$. (In partial cases we have explicit expressions for
$f_{k,\ell}$ from \cite{DDM} but we shall not need them here.)

Let us denote by ~$C^\D$~ the spaces spanned by the elements
~$f_{k,\ell}$, and by ~$L^\D$~ the irreducible subspace of $C^\D$.
 Now in \cite{DDM} was shown:

\nt
{\bf Theorem 2.}~ The irreducible spaces ~$L^\D$~ give a realization
of the irreducible lowest weight representations of ~$\sc(1)$~
given in Theorem 1. \hfill $\dia$

We consider now in more detail the most interesting cases
of the representations ~$L^{(3-p)/2}$~ with $M\neq 0$ and $p\in 2\bbn$.

We first introduce an operator by the polynomial ~$G^2 -2MK \in U(\sc^+)$~
expressed this polynomial in the vector--field realization:
\beq\label{schr}
S ~\doteq~ G^2 -2M K ~~=~~
t^2 \left(\pd^2_x -2M\pd_t \right)
~+~ 2Mt(\half-\D)  \eeq

In these case we have \cite{DDM}:

\nt
{\bf Proposition 2.}~ Each basis polynomial ~$f_{k,\ell}$~ of
$L^{(3-p)/2}$ satisfies:
\bea\label{schrr} S^{p/2}\ f_{k,\ell} ~&=&~
\left( t^2 (\pd^2_x - 2M\pd_t) ~+~ (p-2)Mt
\right)^{p/2} ~ f_{k,\ell} ~=~ \\
~&=&~
t^p~\left( \pd^2_x - 2M\pd_t\right)^{p/2} ~ f_{k,\ell} ~=~ 0 \,, \quad \D = {3-p\over 2}\ . \qquad\qquad\qquad \dia \nn\eea

Thus we have obtained in \eqref{schrr}\
an infinite hierarchy of PDO's ~$S^{p/2}$~ which  give rise to
differential equations
we call   ~{\bf free generalized heat/Schr\"odinger equations}.
The equations are obtained by substituting
the vector field realization in the singular vectors
(thus extending the procedure of \cite{Dob}).
This substitution gives:
\eqnn{cld}
&S^{p/2}
~=~ \left( G^2 - 2MK \right)^{p/2}
~=\cr
&=~ \left( t^2 (\pd^2_x - 2M\pd_t) ~+~ (p-2)mt
\right)^{p/2} ~=\cr
&=~ t^p ~ \left( \pd^2_x - 2M\pd_t\right)^{p/2} \eea
Thus, the hierarchy of equations is:
  \beq\label{hes} t^p ~
\left( \pd^2_x - 2M\pd_t\right)^{p/2} ~f ~=~ 0 \eeq In the case of
function spaces with elements which are polynomials in ~$t$~ (as our
representation spaces) or singular at most as $t^{-p/2}$ for $t\to
0$, the hierarchy is: \eqn{het} \left( \pd^2_x -
2M\pd_t\right)^{p/2} ~f ~=~ 0 \eeq

The above Proposition also shows that
the representation spaces are comprised from solutions
of the corresponding equations \eqref{het}.
The case ~$ p=2 $~ and ~$M$~ real is the
ordinary heat or diffusion equation and for ~$ p = 2 $~ and ~$M$~
purely imaginary we get the free Schr\"odinger equation.
So the members of the hierarchy of equations which are invariant
under the Schr\"odinger group have generically higher orders of
derivatives in ~$t$. This shows that the Schr\"odinger symmetry is not
necessarily connected with first order (in $t$) differential operators.

We can further extend \cite{Dob} to the non-semisimple situation by
considering equations with non-zero RHS. However, invariance w.r.t.
the \S\ requires that the RHS is an element of the irreducible
representation space ~$C^{(p+3)/2}$, while the functions in the LHS
are not restricted to the solution subspace of \eqref{het}. Thus,
using the operator in \eqref{hes} we obtained the following
hierarchy of    generalized heat/Schr\"odinger equations~: \eqn{heu}
t^p ~ \left( \pd^2_x - 2M\pd_t\right)^{p/2} ~f ~=~ j ~, \qquad
f\in\ C^{(3-p)/2} ~, \quad  j\in C^{(3+p)/2} \eeq

\bigskip

\nt {\bf Remark:}~ It is interesting to note that \eqref{heu} looks
similar to an hierarchy of equations involving the d'Alembert
operator and conditionally invariant w.r.t. conformal algebra
$su(2,2)$~: \eqn{hda}
 {\Box}^n ~\hp({\bf x}) ~=~ \hp'({\bf x}) ~, \qquad n\in\bbn \eeq
where ~$\hp,\hp'$~ are scalar fields of different fixed conformal
weights depending on $n$, ~${\bf x} ~=~ (x_0,x_1,x_2,x_3)$~
denotes the Minkowski space-time coordinates, and
~$\Box$~ is the d'Alembert operator:
~$\Box ~=~ \pd^\mu \pd_\mu ~=~ (\vec\pd)^2 - (\pd_0)^2$,
cf., e.g., \cite{Docn}.~$\diamondsuit$

Of course, we may consider more general function
spaces of the variables ~$t,x$,
say ~$C^\infty(\bbr^2,\bbr)$,  on which the
centrally extended \S\ is acting
by formulae \eqref{vecb}. For the example of the ordinary heat equation
one may find also solutions of the type:
\eqn{hea} \left( \a \cosh (\sqrt{\l} x) + \b \sinh (\sqrt{\l} x)
\right) ~\exp\left({ \l t \over 2M}\right) \,, \quad \a,\b,\l\in\bbr \eeq
which results from separation of the ~$t$~ and ~$x$~ variables,
while the polynomial solutions above may be obtained also by
separation of the variables ~$t$~ and ~$Mx^2/2t\,$, cf. \cite{DDM}.

\subsection{Generalized Schr\"odinger equations in the bulk}

In this subsection we review \cite{Dobtre}.
Now we shall employ the bulk vector--field representation \eqref{schr-bulk-one}
trying similarly to the previous subsection to construct generalized Schr\"odinger equations in the bulk.
We start with the operator (distinguishing bulk operators by hats):
\eqn{schr-bu}
\hat{S} ~\doteq~ \hat{G}^2 -2 \pd_- \hat{K} ~=~
t^2 \left(\pd^2_x -2\pd_-\pd_t \right)
~+~ 2t(\half-z\pd_z)\pd_-  ~-~ z^2 \pd_-^2 \eeq
We could use the one-point invariant variable obtained from $u$ by setting in \eqref{udef}
~$t'=0$, $x'=0$, $x'_-=0$, $z'=1$, i.e., we use
\eqn{utdef}
\tu =\frac{4 z}{ x^2 - 2tx_-  +   (z+1)^2} \eeq
Substituting this change in \eqref{schr-bu} we obtain:
\eqn{schr-but}
\hat{S} ~=~ \frac{t^2}{z}\left( \tu^3\del{\tu} + \frac{\tu^4}{2} \del{\tu}^2 \right)\ . \eeq
We shall elaborate on the use of \eqref{schr-but} elsewhere.

Now we set an Ansatz for the fields in the bulk: $ \phi(t,x,x_-,z) =
e^{M x_-} \phi(t,x,z) $ which leads to the identification $ \del{x_-}
= M $  both in the bulk and on the boundary. Thus, we shall use:
\eqn{schr-bus} \hat{S}_0 ~\doteq~ \hat{G}_0^2 -2 M \hat{K}_0 ~=~
t^2 \left(\pd^2_x -2M \pd_t \right) ~+~ 2tM(\half-z\pd_z)  ~-~ z^2
M^2  \eeq

Thus, we obtain the following ~{\it Schr\"odinger-like equation}~ in the bulk:
\eqn{heub}
 \hat{S}_0~\phi ~=~ \phi'
~, \qquad \phi\in\  \hat{C}^{1/2} ~, \quad  \phi'\in \hat{C}^{5/2} \ .\eeq
The relation to the Schr\"odinger equation on the boundary is seen by the following
commutative diagram:
\eqn{diag} \begin{matrix} \hat{C}^{1/2} & {\lra\atop \hat{S}_0} & \hat{C}^{5/2} \cr
&&\cr
\downarrow L_{1/2} & & \downarrow L_{1/2} \cr
&&\cr
C^{1/2} & {\lra \atop S} & C^{5/2} \cr
\end{matrix}
\eeq
where $L_{1/2}$ is the bulk-to-boundary operator defined in \eqref{Buboug}, and \eqref{diag} may be re-written
as the intertwining relation:
\eqn{intr} S \circ L_{1/2} =  L_{1/2} \circ \hat{S}_0 \ ,
\quad {\rm acting~as~operator}\quad ~\hat{C}^{1/2} ~\lra
~ C^{5/2} \eeq
The relation \eqref{intr} (and so \eqref{diag}) follows by substitution of the definitions.

As expected, we have a ~{\it Schr\"odinger-like hierarchy of equations}~ in the bulk:
\eqn{heuz}
 (\hat{S}_0)^{p/2}~\phi ~=~ \phi'
~, \qquad \phi\in\  \hat{C}^{(3-p)/2} ~, \quad  \phi'\in \hat{C}^{(3+p)/2} \ , \quad p\in 2\bbn \eeq

They are equivalent to the Schr\"odinger hierarchy of equations on the boundary \eqref{heu} which is proved by showing
the analogues of \eqref{diag} and \eqref{intr}:
\eqn{diagz} \begin{matrix}
\hat{C}^{\D} & {\lra\atop (\hat{S}_0)^{p/2}} & \hat{C}^{3-\D} \cr
&&\cr
\downarrow L_{\D} & & \downarrow L_{\D} \cr
&&\cr
C^{\D} & {\lra \atop S^{p/2}} & C^{3-\D} \cr
\end{matrix}
\eeq \bea\label{intrz} S^{p/2} \circ L_{\D} &=&  L_{\D} \circ
(\hat{S}_0)^{p/2} \ , \quad {\rm acting~as~operator}\quad
\hat{C}^{\D} ~\lra
~ C^{3-\D}  ,\\
&& \D = (3-p)/2 , \qquad p\in 2\bbn \nn\eea

\vspace{10mm}

\np
\setcounter{equation}{0}
\section{Non-relativistic invariant differential equations for arbitrary
~$n$}

In this Section we review the papers \cite{ADDS,DoSt}.

\subsection{Gauss decomposition of the Schr\"odinger group}

\subsubsection{Triangular decomposition of  ~$\hs(n)$~ for  ~ $n = 2N$}

  \bea
 \hs(2N)& = &  \hs(2N)^+ \oplus \hs(2N)^0 \oplus \hs(2N)^-, \nn \\
\hs(2N)^+ & = & {\rm l.s.} \{ \ G_a,\ K, \ E_{\ell_i \pm \ell_j} \ \}, \label{tri2N} \\
\hs(2N)^0 & = & {\rm l.s.} \{ \ D,\ M, \ H_k,\ \},     \nn \\
\hs(2N)^- & = & {\rm l.s.} \{ \ P_a,\ P_t,\ E_{-(\ell_i \pm
\ell_j)} \ \}, \nn \eea where {\rm l.s.} stands for linear span, $
a, k, i, j $ are integers of $ 1 \leq a \leq 2N,\; 1 \leq k \leq N,
\; 1 \leq i < j \leq N$. For $N>1$ the  generators $ \{ H_k \},\ \{
E_{\ell_i \pm \ell_j} \}, \ \{ E_{-(\ell_i \pm \ell_j)} \} $ are the
Cartan subalgebra generators, the positive root vectors and the
negative root vectors of $so(2N)$, respectively. They are related to
the antisymmetric generators $ \{ J_{ij} \} $ as follows (only the
first line remaining for $N=1$)\bea
    H_k &=& -iJ_{k\; N+k}, \quad (k = 1, 2, \cdots N) \label{2nC-A} \\
    E_{\pm(\ell_j+\ell_k)} &=& -\ha  (J_{jk} \mp iJ_{j\; N+k} \pm iJ_{k\; N+j} -
        J_{N+j\; N+k}), \nn\\ && (1 \leq j < k \leq N) \nn \\
   E_{\ell_j - \ell_k} &=& -\ha (J_{jk} + iJ_{j\; N+k} + iJ_{k\; N+j} +
       J_{N+j\; N+k}), \nn\\ && (1 \leq j \neq k \leq N) \nn
\eea

\subsubsection{Triangular decomposition of  ~$\hs(n)$~ for  ~ $n = 2N+1$}

  \bea
 \hs(2N+1) & = & \hs(2N+1)^+ \oplus\hs(2N+1)^0\oplus\hs(2N+1)^-, \nn \\
 \hs(2N+1)^+ & = & {\rm l.s.} \{ \ G_a,\ K, \ E_{\ell_k},\ E_{\ell_i \pm \ell_j} \ \}, \label{tri2N1} \\
 \hs(2N+1)^0 & = & {\rm l.s.} \{ \ D,\ M, \ H_k,\ \},     \nn \\
 \hs(2N+1)^- & = & {\rm l.s.} \{ \ P_a,\ P_t,\ E_{-\ell_k},\ E_{-(\ell_i \pm \ell_j)} \ \}, \nn
\eea where $ a, k, i, j $ are integers of $ 1 \leq a \leq 2N+1,\; 1
\leq k \leq N, \; 1 \leq i < j \leq N. $ The  generators $ \{ H_k
\},\ \{ E_{\ell_k},\ E_{\ell_i \pm \ell_j} \}, \ \{ E_{-\ell_k},\
E_{-(\ell_i \pm \ell_j)} \} $ are the Cartan subalgebra generators,
the positive root vectors and the negative root vectors of
$so(2N+1)$, respectively. They are related to the antisymmetric
generators $ \{ J_{ij} \} $ as in the case $\ n=2N\ $ for  $ H_k\,$,
$E_{\pm(\ell_j+\ell_k)}\,$,
 and $ E_{\ell_j - \ell_k} $, while $ E_{\pm \ell_k}$ is defined
 by:\beq
  E_{\pm \ell_k} = -\frac{1}{\sqrt 2}(J_{k\; 2N+1} \mp i J_{N+k\; 2N+1}),
      \quad (k = 1, 2, \cdots, N) \label{2n1C-A}
   \eeq

\subsubsection{Gauss decomposition of the Schr\"odinger group ~$\hat{\bf S}(n)$}
Let ~$ g \in \hat{\bf S} (n) $~ be an element with Gauss decomposition: ~$ g
= g_+\, g_0\, g_-\, $.

\noindent For even ~$ n = 2N $~ we have: \bea
  & & g_+ = \left(  \prod_{a=1}^{2N} e^{x_a G_a}\right) e^{tK}
      \left( \prod_{1\leq j< k \leq N} e^{\xi_{jk} E_{\ell_j+\ell_k}} \right)
      \left( \prod_{1\leq j< k \leq N} e^{\eta_{jk} E_{\ell_j-\ell_k}}\right)
      \nn \\
  & & g_0 = e^{\delta D} e^{m M} \prod_{k=1}^N e^{h_k H_k}, \label{gauss2n} \\
  & & g_- = \left(  \prod_{a=1}^{2N} e^{u_a P_a}\right) e^{\kappa P_t}
      \left( \prod_{1\leq j< k \leq N} e^{\xi^-_{jk} E_{-\ell_j-\ell_k}} \right)
      \left( \prod_{1\leq j< k \leq N} e^{\eta^-_{jk} E_{-\ell_j+\ell_k}}\right)
      \nn
\eea

\bigskip

\noindent For odd ~$ n = 2N +1 $~ we have: \bea
    g_+ &=& \left(  \prod_{a=1}^{2N+1} e^{x_a G_a}\right) e^{tK}
      \left( \prod_{k=1}^N e^{\varphi_k E_{\ell_k}} \right)\nn\\
&&\times\ \left( \prod_{1\leq j< k \leq N} e^{\xi_{jk}
E_{\ell_j+\ell_k}} \right)
      \left( \prod_{1\leq j< k \leq N} e^{\eta_{jk} E_{\ell_j-\ell_k}}\right)
      \nn \\
   g_0 &=& e^{\delta D} e^{m M} \prod_{k=1}^N e^{h_k H_k}  \label{gauss2n1} \\
   g_- &=& \left(  \prod_{a=1}^{2N+1} e^{u_a P_a}\right) e^{\kappa P_t}
      \left( \prod_{k=1}^N e^{\varphi^-_k E_{-\ell_k}} \right)\nn\\
&&\times\      \left( \prod_{1\leq j< k \leq N} e^{\xi^-_{jk}
E_{-\ell_j-\ell_k}} \right)
      \left( \prod_{1\leq j< k \leq N} e^{\eta^-_{jk} E_{-\ell_j+\ell_k}}\right)
      \nn
\eea

\subsection{Representations of ~$ \hat{S}(n)$}
Let briefly recall the SSLG setting \cite{Dob} adapted to the Scr\"odinger setting above:\\
\bu ~${\cal C}_{\Lambda}$: the space of ${\bbc}^{\infty}$ functions ${\cal
F}$ on $\hat{\bf S} (n)$  with right covariance  property \beq {\cal F}(g
x g')=e^{\Lambda(H)}{\cal F}(g),\eeq where ~$x=e^H\in g_0\,$, ~$H\in
\hat{s}^0(n)$, ~$g'\in g_-\,$, ~$\Lambda \in (\hat{s}^0(n))^*$. We
 use the following notation for the values of $\Lambda(H)$:
~$\L(M)=-m$, ~$\L(D)=\D$, ~$\L(H_k)=-h_k\in\bbz$.\\ \bu
~$C_{\Lambda}$: space of restricted functions $\psi ={\cal
F}\left.\right|_{g_+}$\ .\\ \bu ~$\pi_L(X)$: left action of $X \in
\hat{s}(n)$ on ${\cc}_{\Lambda}$ \beq \pi_L(X){\cal
F}(g)=\frac{d}{d\tau}\left. {\cal F}(e^{-\tau
X}g)\right|_{\tau=0}\label{eq:leftact} \eeq \bu ~$\pi_R(X)$: right
action of $X \in \hat{s}(n)$ on ${\cc}_{\Lambda}$\beq \pi_R(X){\cal
F}(g)=\frac{d}{d\tau}\left. {\cal F}(ge^{\tau
X})\right|_{\tau=0}\label{eq:rightact} \eeq

\subsubsection{Representations for ~$n=2N$}
Define \beq
  G_{k}^{\pm} = G_k \pm i G_{N+k}, \qquad x_k^{\pm} = \ha (x_k \mp i x_{N+k}),
  \qquad k = 1, 2, \cdots, N   \label{Gpm}
\eeq then \beq
  \prod_{a=1}^{2N} e^{x_a G_a} = \prod_{k=1}^N e^{x_k^+ G_k^+} e^{x_k^- G_k^-}. \label{GpGm}
\eeq Sometimes the use of (\ref{GpGm}) is more convenient. We denote
each factor of $ g_+ $ by \bea
   g_+ &=& \Gamma_x e^{tK} \Gamma_{\xi} \Gamma_{\eta}, \label{Gamma} \\
    \Gamma_x &=&  \prod_{a=1}^{2N} e^{x_a G_a}, \quad
      \Gamma_{\xi} =  \prod_{1\leq j< k \leq N} e^{\xi_{jk} E_{\ell_j+\ell_k}},
      \quad \Gamma_{\eta} = \prod_{1\leq j< k \leq N} e^{\eta_{jk} E_{\ell_j-\ell_k}} \nn
\eea

\bigskip

\nt
{\bf Action of ~\boldmath{$\hs(2N)^0$~:}}\\
Following the general procedure of \cite{Dob} and results of
\cite{ADDS,DoSt} we obtain the following formulae  for the vector-field
representation: \bea
 & & \pi_L(M) = -\Lambda(M),
\qquad \pi_L(D) = -\Lambda(D) - \sum_{a=1}^{n} x_a \del{x_a}-2t\del{t} \label{MDeven} \\
 & & \pi_L(H_k) = -\Lambda(H_k) + x_k^+ \del{x_k^+} - x_k^- \del{x_k^-} \nn \\
 & & \quad
     - \sum_{i=1}^{k-1} \left( \xi_{ik}\del{\xi_{ik}} - \eta_{ik} \del{\eta_{ik}} \right)
     - \sum_{j=k+1}^N \left( \xi_{kj} \del{\xi_{kj}} + \eta_{kj} \del{\eta_{kj}} \right)
     \label{Hkeven}\eea The $x$-dependent parts of (\ref{Hkeven}) can be rewritten as:
\beq
  i \left( x_k \del{x_{N+k}}  - x_{N+k} \del{x_k}  \right)
  \label{xparteven}
\eeq

\bigskip

\nt {\bf Action of ~\boldmath{$\hs(2N)^+$}}~ \cite{ADDS}~:\\
\bea
 \pi_L(G_a) &=& -\del{x_a} \label{Geven} \ , \qquad \pi_L(K) = -\del{t} \label{GKeven} \\
   \pi_L(E_{\ell_j+\ell_k}) &=& x_j^+ \del{x_k^-} - x_k^+ \del{x_j^-} - \del{\xi_{jk}}
      \label{Epeven} \\
 \pi_L(E_{\ell_j-\ell_k}) &=& x_j^+ \del{x_k^+} - x_k^- \del{x_j^-}
      - \sum_{s=k+1}^N \xi_{ks} \del{\xi_{js}} - \sum_{r=1}^{j-1} \xi_{rk} \del{\xi_{rj}} \nn \\
  & &
      +\ \sum_{r=j+1}^{k-1} \xi_{rk} \del{\xi_{jr}} + \sum_{r=1}^{j-1} \eta_{rj} \del{\eta_{rk}}
      - \del{\eta_{jk}}       \label{Emeven}
\eea

\bigskip

\nt {\bf Action of ~\boldmath{$\hs(2N)^-$}~:} ~Standardly, the
left action is determined from the formula \beq e^{-\tau X}g_+g_0g_-
= g'_+g'_0g'_-,\eeq where $X\in g_-$. For us, it will be enough to
find $g'_+$ and $g'_0$ because the vector field representation is
restricted to $g_+$.

We introduce the generators: ~$P_k^\pm \doteq P_k \pm i P_{N+k}\,$.
We have:
  \bea e^{-\tau P_k^+}g & = & \left( \prod_{r=1}^N e^{x_r'^+ G_r^+} e^{x_r'^- G_r^-} \right)
  e^{t K} \left( \prod_{r<s} e^{\xi_{rs} E_{\ell_r + \ell_s}}
  \right)\left( \prod_{r<s} e^{\eta_{rs} E_{\ell_r - \ell_s}}
  \right)\times\nn\\
  & \times & e^{\delta D}e^{m'M}\left(\prod_{r=1}^N e^{h_rH_r}\right)g'_-\ , \\
&&
 x_k'^+ = x_k^+ - \tau t\,,  \quad
  x_r'^{\pm} = x_r^{\pm} \ ({\rm otherwise}),
\qquad m'= m-2\tau x_k^-\ , \nn \\
 e^{-\tau P_k^-}g & = & \left( \prod_{r=1}^N e^{x_r^+ G_r^+} e^{x_r'^- G_r^-} \right)
  e^{t K} \left( \prod_{r<s} e^{\xi_{rs} E_{\ell_r + \ell_s}}
  \right)\left( \prod_{r<s} e^{\eta_{rs} E_{\ell_r - \ell_s}}
  \right)\times\nn\\
  & \times & e^{\delta D}e^{m'M}\left(\prod_{r=1}^N e^{h_rH_r}\right)g'_-\ , \\
&&   x_k'^- = x_k^- - \tau t\,,  \quad
  x_r'^{-} = x_r^{-} \ ({\rm otherwise}),
  \qquad m'= m-2\tau x_k^+ \ ,\nn \\
 e^{-\tau P_t}g & = & \left( \prod_{r=1}^N e^{x_r'^+
G_r^+} e^{x_r'^- G_r^-} \right)
  e^{t' K} \left( \prod_{r<s} e^{\xi_{rs} E_{\ell_r + \ell_s}}
  \right)\left( \prod_{r<s} e^{\eta_{rs} E_{\ell_r - \ell_s}}
  \right)\times\nn\\
  & \times & e^{\delta' D}e^{m'M}\left(\prod_{r=1}^N e^{h_rH_r}\right)g'_-\ ,\label{eq:lactpt} \\
 & & x_j'^+ = x_j^+ - \tau tx_j^+, \quad x_j'^- = x_j^- -
\tau tx_j^-\,\quad t'=t-\tau t^2\ , \nn\\ & & \delta'=\delta-\tau
t,\quad m'=m-2\tau \sum_{j=1}^N x_j^+x_j^-\ , \nn\eea
    \bea e^{-\tau E_{-(\ell_j+\ell_k)}}g & = & \left(
\prod_{r=1}^N e^{x_r'^+ G_r^+} e^{x_r^- G_r^-} \right)
  e^{t K} \left( \prod_{r<s} e^{\xi_{rs}' E_{\ell_r + \ell_s}}
  \right)\left( \prod_{r<s} e^{\eta_{rs}' E_{\ell_r - \ell_s}}
  \right)\ \times \nn\\ && \times \
   e^{\delta D}e^{mM}\left(\prod_{r=1}^Ne^{{h'}_rH_r}\right)g'_-\ ,  \\
 & & x_i'^+ = x_i^+ - \tau
(\delta_{ij}x_k^--\delta_{ik}x_j^-), \quad x_i'^- = x_i^-\ ,
\nn\\
& & \xi_{rs}' = \xi_{rs}-\delta_{jr}\delta_{ks}\xi_{rs}^2\ ,
~~ \xi_{sr}' = \xi_{sr}-\delta_{js}\delta_{kr}\xi_{sr}^2 \nn\\
& &
\eta_{rs}'=\eta_{rs}-\tau\{\delta_{sk}(\xi_{ir}-\xi_{ri})+\delta_{sj}(\xi_{rk}-\xi_{kr})\}\ , \nn\\
&&{h'}_l=h_l-\tau(\delta_{jl}+\delta_{kl})\xi_{jk} \ ,\nn \eea \bea
e^{-\tau E_{-(\ell_j-\ell_k)}}g & = & \left( \prod_{r=1}^N e^{x_r'^+
G_r^+} e^{x_r'^- G_r^-} \right)
  e^{t K} \left( \prod_{r<s} e^{\xi_{rs}' E_{\ell_r + \ell_s}}
  \right)\left( \prod_{r<s} e^{\eta_{rs}' E_{\ell_r - \ell_s}}
  \right)\ \times\nn\\ &&\times \ e^{\delta D}e^{mM}\left(\prod_{r=1}^Ne^{{h'}_rH_r}\right)g'_-\ , \\
 & & x_i'^+ = x_i^+ + \tau
\delta_{ij}x_k^+, \quad x_i'^- = x_i^--\tau\delta_{ik}x_j^-\ ,
\nn\\
& & \xi_{rs}' =
\xi_{rs}-\tau\delta_{kr}\xi_{is}-\tau\delta_{sk}\xi_{ri}\ ,
\quad {h'}_l=h_l-\tau(\delta_{kl}-\delta_{jl})\xi_{jk} \ ,  \nn\\
& &
\eta_{rs}'=\eta_{rs}-\tau\delta_{rk}\eta_{js}+\tau\delta_{js}\eta_{rk}-\delta_{rj}\delta_{sk}\eta_{rs}^2
\ .\nn  \eea The above formulae are correct up to terms quadratic in
$\tau $. This is enough since these formulae  are used only to
obtain the left  action of the corresponding generators of the
algebra  (see formulae \eqref{eq:leftact}).

 Thus, from the infinitesimal left action
 we obtain the  vector-field representation:
    \bea \pi_L(P_k^+) & = & -t\del{x_k^+}-2\Lambda(M)x_k^- \ ,\\
\pi_L(P_k^-) & = & -t\del{x_k^-}-2\Lambda(M)x_k^+\ , \nn\\
\pi_L(P_t) & = &
-t^2\del{t}-t\sum_{j=1}^N\left(x_j^+\del{x_j^+}+x_j^-\del{x_j^-}\right)
-2\Lambda(M)\sum_{j=1}^Nx_j^+x_j^--\Lambda(D)t\ ,\nn\\
\pi_L(E_{-(\ell_j+\ell_k)}) & = &
-(\Lambda(H_j)+\Lambda(H_k))\xi_{jk}+
x_j^+\del{x_k^-}-x_k^+\del{x_j^-}-\xi_{ik}^2\del{\xi_{ik}}\ - \nn\\
& &-\
\sum_{p=j+1}^{k-1}\xi_{jp}\del{\eta_{pk}}-\sum_{p=1}^{j-1}\xi_{pk}\del{\eta_{pj}}
+\sum_{p=1}^{j-1}\xi_{pj}\del{\eta_{pk}}\ ,\nn\\
\pi_L(E_{-(\ell_j-\ell_k)}) & = &
-(\Lambda(H_k)-\Lambda(H_j))\xi_{jk}+
x_k^+\del{x_j^+}-x_j^-\del{x_k^-}+\eta_{jk}^2\del{\eta_{jk}}\ -\nn\\
&& -\ \sum_{s=j+1}^{k-1}\xi_{js}\del{\xi_{sk}}+\sum_{s=k+1}^N\xi_{js}\del{\xi_{ks}}\ -\nn\\
&& -\
\sum_{r=1}^{j-1}\xi_{rj}\del{\xi_{rk}}-\sum_{q=k+1}^N\eta_{jq}\del{\eta_{kq}}\nn
\eea

\subsubsection{Representations  for ~$n=2N+1$}

In addition to $ G_k^{\pm}, x_k^{\pm} $ given by (\ref{Gpm}), let us
introduce \beq
  G_0 = \sqrt{2} G_{2N+1}, \qquad x_0 = \frac{1}{\sqrt 2} x_{2N+1}, \label{G0}
\eeq then \beq
  \prod_{a=1}^{2N+1} e^{x_a G_a} = e^{x_0 G_0} \prod_{k=1}^N
  e^{x_k^+ G_k^+} e^{x_k^- G_k^-}. \label{G0GpGm}
\eeq The difference from $ n=2N $ is the existence of $ G_0 $ and $
E_{\ell_k}$.

\bigskip

\nt
{\bf Action of ~\boldmath{$\hs(2N+1)^0$}~:}\\
Since $ [D, G_0 ] = D_0, \ [D, E_{\ell_k}] = 0 $, the left
representation of $M, D $ are the same as in the $n$-even case - cf.
(\ref{MDeven}), while for $H_k$ we have:   \beq
 \pi_L(H_k) = \pi_L(H_k)^{\rm even} ~-~ \varphi_k \del{\varphi_k}
     \label{Hkodd}
  \eeq
where ~$\pi_L(H_k)^{\rm even}$~ is the RHS of formula
(\ref{Hkeven}).

\bigskip

\nt
{\bf Action of ~\boldmath{$\hs(2N+1)^+$}~:}\\
The representations of $ G_a,\ K,\ E_{\ell_j + \ell_k} $ are the
same as in the $n$-even case - cf. (\ref{GKeven}) with $ a = 1,2,
\cdots, n=2N+1 $, while for the others we have \cite{ADDS}:
 \beq
  \pi_L(E_{\ell_k}) = x_k^+ \del{x_0} - x_0
  \del{x_k^-}-\del{\varphi_k}
  - \sum_{i=1}^{k-1} \varphi_i \del{\xi_{ik}} + \sum_{j=k+1}^N \varphi_j \del{\xi_{kj}}
  \label{Eodd}
\eeq   \beq
    \pi_L(E_{\ell_j-\ell_k}) = \pi_L(E_{\ell_j-\ell_k})^{\rm even} ~
      - ~\varphi_k \del{\varphi_j}      \label{Emodd}
\eeq where ~$\pi_L(E_{\ell_j-\ell_k})^{\rm even}$~ is the RHS of
(\ref{Emeven}).

\bigskip

\nt {\bf Action of ~\boldmath{$\hs(2N+1)^-$}~:}\\
Since $P_k^{\pm}$ commute with $G_0$ and $[P_k^+, E{\ell_k}]=P_0$
gives rise only in changing of $g_-$, the left action of $P_k^{\pm}$
is the same as in the even case. Next, \beq  [P_t, G_0^n]= n
G_0^{n-1}+n(n-1)G_0^{n-2}M \ , ~~
 [P_t, E_{\ell_k}]= -P_k^- \eeq provide only \beq
m'=m'_{even}-\tau x_0^2.\nn\eeq from which follows \beq
\pi_L(P_t)=\pi_L(P_t)^{even}-\Lambda(M)x_0^2\ .\eeq For the left
action of $E_{-(\ell_j-\ell_k)}$ the unique non-vanishing commutator
is \beq [E_{-(\ell_j-\ell_k)}, E_{\ell_k}]=E_{\ell_k} \nn\eeq which
means that \beq \varphi_z'=\varphi_z-\tau\delta_{zk}\varphi_j\nn\eeq
and consequently\beq
\pi_L(E_{-(\ell_j-\ell_k)})=\pi_L(E_{-(\ell_j-\ell_k)})^{even}-\varphi_j\del{\varphi_k}.
\eeq

For the rest of the generators we first obtain:
  \bea e^{-\tau P_0}g  &=&  e^{x'_0G_0}\left(
\prod_{r=1}^N e^{x_r^+ G_r^+} e^{x_r^- G_r^-} \right)
  e^{t K} \left( \prod_{r<s} e^{\xi_{rs} E_{\ell_r + \ell_s}}
  \right)\left( \prod_{r<s} e^{\eta_{rs} E_{\ell_r - \ell_s}}
  \right)\ \times \nn\\ &&\times\ e^{\delta D}e^{m'M}\prod_{r=1}^Ne^{h_rH_r}g'_-\ ,\\
&& x'_0 = x_0-\tau t, \qquad m' = m - 2\tau x_0\ ,\nn\eea
 \bea e^{-\tau E_{-\ell_k}}g & = & e^{x'_0G_0}\left(
\prod_{r=1}^N e^{{x'}_r^+ G_r^+} e^{x_r^- G_r^-} \right)
  e^{t K} \left(\prod_{r=1}^Ne^{\varphi'_r E_{\ell_r}}\right)\ \times\\
  && \times\  \left( \prod_{r<s} e^{\xi_{rs} E_{\ell_r + \ell_s}}
  \right)\left( \prod_{r<s} e^{\eta'_{rs} E_{\ell_r - \ell_s}}
  \right)e^{\delta D}e^{mM}\left(\prod_{r=1}^Ne^{h'_rH_r}\right)g'_-\ ,\nn\\
&& x'_0=x_0+\tau x_k^-\ ,\qquad
{x'}_r^+=x_r^+-\tau\delta_{rk}x_0\ ,\nn\\
&&
\varphi'_z=\varphi_z-\tau\delta_{zk}\frac{\varphi_k^2}{2}+\tau(\delta_{zs}\xi_{ks}+\delta_{zr}
\xi_{rk}-\delta_{qz}\eta_{kq})\ ,\nn\\
&& \eta'_{pq}=\eta_{pq}-\tau\delta_{zp}\delta_{kq}\varphi_z, \qquad
h_z'=h_z-\tau\delta_{kz}\varphi_z \ ,\nn\eea \bea e^{-\tau
E_{-(\ell_j+\ell_k)}}g & = & e^{x_0G_0}\left( \prod_{r=1}^N
e^{{x'}_r^+ G_r^+} e^{{x'}_r^- G_r^-} \right)
  e^{t K} \left(\prod_{r=1}^Ne^{\varphi'_r E_{\ell_r}}\right)\ \times \\
  && \times \ \left( \prod_{r<s} e^{\xi_{rs}' E_{\ell_r + \ell_s}}
  \right)\left( \prod_{r<s} e^{\eta'_{rs} E_{\ell_r - \ell_s}}
  \right)e^{\delta D}e^{mM}\left(\prod_{r=1}^Ne^{h'_rH_r}\right)g'_-\ ,\nn\\
 && {x'}_r^+ = ({x'}_r^+)^{even},\qquad {x'}_r^- =
({x'}_r^-)^{even}\ ,\nn\\
&& {\xi_{rs}}'= ({\xi_{rs}}')^{even},\qquad {\eta_{rs}}'=
({\eta_{rs}}')^{even},\qquad {h_l}'= ({h_l}')^{even}\ ,\nn\\
&& \varphi'_z=
\varphi_z-\tau(\delta_{jz}\varphi_z\xi_{zk}-\delta_{kz}\varphi_z\xi_{jz})\
.\nn\eea

From the above follow the vector-field representation:
\bea \pi_L(P_0) &=&  -\,t\del{x_0}-2x_0\Lambda(M)\ , \\
 \pi_L(E_{-\ell_k}) & = &
-\,\varphi_k\Lambda(H_k)+x_k^-\del{x_0}-x_0\del{x_k^-}\ -\nn\\
&  &-\ \frac{\varphi_k^2}{2}\del{\varphi_k}+\sum_{s=k+1}^N\xi_{ks}
\del{\varphi_s}+\sum_{r=1}^{k-1}\xi_{rk}\del{\varphi_r}\ -\nn\\
& &-\
\sum_{q=k+1}^N\eta_{kq}\del{\varphi_q}-\sum_{p=1}^{k-1}\varphi_p\del{\eta_{pk}}\ ,\\
\pi_L(E_{-(\ell_j+\ell_k)})&=&\pi_L(E_{-(\ell_j+\ell_k)})^{even}-\xi_{jk}(\varphi_j\del{\varphi_j}
-\varphi_k\del{\varphi_k})\eea

\subsection{Singular vectors and invariant equations for ~$\hs(2N)$}

\subsubsection{Singular vectors}

  Let $ v_0 $ be the lowest weight vector of the Verma module $V^\L$  with
lowest weight $ \Lambda\,$: \beq
  H v_0 = \Lambda(H) v_0\,, \qquad X v_0 = 0, \quad {\rm for} \quad
 \forall\, H \in {\mathfrak g}^0, ~ \forall\, X \in {\mathfrak g}^-
  \label{LWV} \eeq
A singular vector is a vector $v_s$ in the Verma module $V^\L$ such
that: \beq H v_s = \Lambda'(H) v_s\,, \qquad X v_s = 0, \quad {\rm
for} \quad
 \forall\, H \in {\mathfrak g}^0, ~ \forall\, X \in {\mathfrak g}^-\ ,
 ~~\L'\neq \L  \label{eq:defsvec}
\eeq Since the singular vectors are in   $ V^{\Lambda} \simeq
U({\mathfrak g}) v_0 $, let us introduce \cite{ADDS}\bea
 v_{\alpha\beta\gamma\lambda\rho} &=& \left\{ \prod_{r=1}^N (G_r^+)^{\alpha_r} \right\}
    \left\{ \prod_{r=1}^N (G_r^-)^{\beta_r} \right\} K^{\gamma}
    \left\{ \prod_{1\leq r \leq N} (E_{\ell_r+\ell_s})^{\lambda_{rs}} \right\}
\nn\\ &&\times   \left\{ \prod_{1\leq r \leq N}
(E_{\ell_r-\ell_s})^{\rho_{rs}} \right\}
    v_0,
    \label{vabg}
\eea where $ \alpha, \beta, \lambda $ and $ \rho $ are abbreviation
of sets of non-negative integers \bea
 & & \alpha = [ \alpha_1, \alpha_2, \cdots, \alpha_N ], \qquad \qquad \;
     \beta = [ \beta_1, \beta_2, \cdots, \beta_N ] \nn \\
 & & \lambda = [\lambda_{12}, \lambda_{13}, \cdots, \lambda_{N-1\; N} ], \qquad
     \rho = [ \rho_{12}, \rho_{13}, \cdots, \rho_{N-1\;N} ]
     \label{int}
\eea and $ \gamma $ is also a non-negative integer.

  Elementary computation shows that the action of ${\mathfrak g}^0 $ on
$ v_{\alpha\beta\gamma\lambda\rho} $ is given by \bea
  M v_{\alpha\beta\gamma\lambda\rho} &=& \Lambda(M) v_{\alpha\beta\gamma\lambda\rho} =
  -m v_{\alpha\beta\gamma\lambda\rho}
      \nn \\
   D v_{\alpha\beta\gamma\lambda\rho} &=& (\L(D)+p)v_{\alpha\beta\gamma\lambda\rho}
=  (p-d) v_{\alpha\beta\gamma\lambda\rho}
      \nn \\
   H_k\, v_{\alpha\beta\gamma\lambda\rho} &=& (\L(H_k)+r_k) v_{\alpha\beta\gamma\lambda\rho}
=   (r_k - h_k) v_{\alpha\beta\gamma\lambda\rho}
      \label{MDHk2n}\\
&&   p = \sum_{r=1}^N (\alpha_r + \beta_r) + 2\gamma   \label{prk2n} \\
&&   r_k = -\alpha_k + \beta_k
      +\ \sum_{s=k+1}^N (\lambda_{ks} + \rho_{ks})
  \ + \sum_{r=1}^{k-1}(\lambda_{rk}-\rho_{rk})
         \nn\eea

 The singular vectors have the following form of
\beq
  v_s = \sum_{\alpha,\lambda,\rho} a_{\alpha\lambda\rho}
      v_{\alpha\beta\gamma\lambda\rho}
  \label{svform}
\eeq where we fix the values of $ p, r_k\ (k=1,2,\cdots, N) $ to
make $ v_{\alpha\beta\gamma\lambda\rho} $ homogeneous. Let us look
for the singular vectors with non-zero central element: $ \Lambda(M)
\neq 0. $

Applying ~$Xv_s=0$~ \eqref{eq:defsvec} for ~$X=P_k^\pm,$, ~$X=P_t$,  $X= E_{-(\ell_i \pm \ell_j)} v_s \ (i<j)$,
in \cite{DoSt} was obtained  the
general explicit formula for the singular vectors: \beq v_s=c
\left( \sum_{a=1}^{2N} G_a^2 -2\Lambda(M) K \right)^{p/2}
\sum_{\lambda,\rho}\left(\prod_{r<s}E_{\ell_r+\ell_s}^{\lambda_{rs}}
\prod_{p<q}E_{\ell_p-\ell_q}^{\rho_{pq}}\right)v_0\ .\label{eq:gsv2N}\eeq

\bigskip

\nt
The ~{\bf Example  N=1}~ is trivial since \eqref{eq:gsv2N} reduces just to:
\beq v_s=c
\left( \sum_{a=1}^{2} G_a^2 -2\Lambda(M) K \right)^{p/2}\, v_0
 \ .\label{eq:gsv2}\eeq

\bigskip

\nt
{\bf Example: N=2}\\
We have ~$i=1, j=2$, ~$r_1=\lambda_{12}+\rho_{12},\quad
r_2=\lambda_{12}-\rho_{12}$ ~are fixed, and \beq
\lambda_{12}=(1+h_1+h_2)/3,\qquad \rho_{12}=(h_1-h_2-1)/3.\nn\eeq
Consequently the form of the singular vectors is:   \beq v_s=c
\left( \sum_{a=1}^4G_a^2-2\Lambda(M) K \right)^{p/2}\left( 1+
E_{\ell_1+\ell_2}^{\lambda_{12}}+E_{\ell_1-\ell_2}^{\rho_{12}}+
E_{\ell_1+\ell_2}^{\lambda_{12}}E_{\ell_1-\ell_2}^{\rho_{12}}\right)v_0\
.\eeq

\bigskip

\nt{\bf Example: N=3}\\
{\bf 1:}~ $i=1, j=2$\\
From the extra conditions listed above, it follows: $\lambda_{13}=0,
\rho_{13}=0.$\\
{\bf 2:}~$i=1, j=2$\\
In this case we have $\lambda_{12}=0.$\\
{\bf 3:}~ $i=2, j=3$\\
From the extra conditions listed above, it follows: $\lambda_{12}=0,
\rho_{13}=0.$\\  This leads to\beq r_1 =  \rho_{12}\ , \quad r_2  =
\lambda_{23}+\rho_{23}-\rho_{12}\ , \quad r_3  =
\lambda_{23}-\rho_{23} \eeq On the other hand\beq \rho_{12}  =
1+h_1-h_2+r_2-r_1\ , \quad \lambda_{23}  =  1+h_2+h_3-r_2-r_3\
,\quad \rho_{23}  =  1+h_2-h_3+r_3-r_2 \eeq It follows \beq
\rho_{12}=(3h_1-h_2+5)/7,\quad \rho_{23}=(h_1+2h_2+4)/7-h_3/3,\quad
\lambda_{23}=(h_1+2h_2+4)/7+h_3/3.\nn\eeq The most general form of
the singular vector for ~$n=2N=6$~ is: \beq v_s=c \left(
\sum_{a=1}^6G_a^2-2\Lambda(M) K
\right)^{p/2}\sum_{q=0,\lambda_{23}\,;\ r=0,\rho_{12}\,;\
s=0,\rho_{23}}E_{\ell_2+\ell_3}^q
E_{\ell_1-\ell_2}^rE_{\ell_2-\ell_3}^s v_0\ .\eeq

\subsubsection{Invariant equations}

Following the general procedure of \cite{Dob} to obtain the
invariant operators and equations we must substitute any generator
$X$ in the expression \eqref{eq:gsv2N} for the singular vector with
the right action ~$\pi_R(X)$~ given in \cite{ADDS}.   Thus, we
obtain
  the invariant equations
\beq \left( \sum_{a=1}^{2N} \frac{\partial^2}{\partial x_a^2} + 2M
\del{t} \right)^{p/2}
\sum_{\lambda,\rho}\left(\prod_{r<s}\pi_R(E_{\ell_r+\ell_s})^{\lambda_{rs}}\prod_{p<q}\pi_R(
 E_{\ell_p-\ell_q})^{\rho_{pq}}\right)\psi ~=~
 \psi' \label{hier2n}\eeq where $\pi_R(E_{\ell_r+\ell_s}), \pi_R(E_{\ell_p-\ell_q})$
 are given by formulae (4.27),(4.28) of \cite{ADDS}, ~$ m = -\Lambda(M) $, and ~$\psi'$ belongs to the
representation ~$\L'$~ such that ~$\L'(M)=\L(M)$,
~$\L'(H_k)=\L(H_k)+r_k$ and ~$\L'(D)=\L(D)+p = \D+p = \ha p +N+1$.

In particular for $ N=2$ one obtains: \beq
 \left( \sum_{a=1}^{4} \frac{\partial^2}{\partial x_a^2} + 2M \del{t} \right)^{p/2}
 \sum_{\lambda=0,\lambda_{12}\,;\ \rho=0,\rho_{12}}\left(\del{\xi_{12}}\right)^\lambda
 \left(\del{\eta_{12}}\right)^\rho\psi=\psi'
 \label{hier2.2}  \eeq while for $N=3$ the invariant equation is
 written in the form\beq
 \left( \sum_{a=1}^{6} \frac{\partial^2}{\partial x_a^2} + 2M \del{t} \right)^{p/2}
 \sum_{q,r,s}\left(\del{\xi_{23}}+\eta_{13}\del{\xi_{12}}\right)^q
 \left(\del{\eta_{12}}+\eta_{23}\del{\eta_{13}}\right)^r\left(\del{\eta_{23}}\right)^s\psi=\psi'
 \label{hier3.2}  \eeq
where $q=0,\lambda_{23}\,;\ r=0,\rho_{12}\,;\ s=0, \rho_{23}\,$.

\subsection{Singular vectors and invariant equations for ~$\hs(2N+1)$}

\subsubsection{Singular vectors}

In comparison to the  $ n=2N$ case we have additional elements $ G_0
$ and $ E_{\ell_k} $. Let us introduce \bea
  v_{\omega\alpha\beta\gamma\sigma\lambda\rho} &=&
     G_0^{\omega} \left\{ \prod_{r=1}^N (G_r^+)^{\alpha_r} \right\}
     \left\{ \prod_{r=1}^N (G_r^-)^{\beta_r} \right\} K^{\gamma}
     \\
  &\times&
    \left\{ \prod_{r=1}^N E_{\ell_r}^{\sigma_r} \right\}
    \left\{ \prod_{1\leq r<s \leq N} (E_{\ell_r+\ell_s})^{\lambda_{rs}} \right\}
    \left\{ \prod_{1\leq r<s \leq N} (E_{\ell_r-\ell_s})^{\rho_{rs}} \right\}
    v_0 \nn
    \label{voab}
\eea where $ \sigma = [\sigma_1, \sigma_2, \cdots, \sigma_N] $ is a
set of non-negative integers. The action of ~$M,D,H_k$~ is given as
in the $n=2N$ case, i.e., by (\ref{MDHk2n}), but the values of $p$
and $r_k$ now are given by:  \bea
  p &=& \omega + \sum_{r=1}^N (\alpha_r + \beta_r) + 2\gamma, \label{prkodd} \\
  r_k &=& -\alpha_k + \beta_k + \sigma_k
      + \sum_{r=k+1}^N (\lambda_{kr} + \rho_{kr})
+ \sum_{r=1}^{k-1} (\lambda_{rk}-\rho_{rk})\nn\eea

The singular vectors have the form of \beq
  v_s = \sum_{\omega,\alpha,\sigma,\lambda,\rho}
        a_{\omega\alpha\sigma\lambda\rho} v_{\omega\alpha\sigma\lambda\rho}.
        \label{vsodd}
\eeq

Imposing ~$Xv_s=0$~ for ~$X=  P_0 \doteq \sqrt 2 P_{2N+1}\,$,
~$X=P_k^\pm\,,P_0\,$, ~$X=E_{-\ell_k}$, ~$X=E_{-(\ell_i\pm\ell_j)}$
in \cite{DoSt} was obtained
  the   general explicit formula for the singular
vectors: \beq v_s=c \left( \sum_{a=1}^{2N+1} G_a^2 -2\Lambda(M) K
\right)^{p/2}
\sum_{\sigma=0,\ell_N;\lambda;\rho}\left(E_{\ell_N}^{\sigma}
\prod_{r<s}E_{\ell_r+\ell_s}^{\lambda_{rs}}
\prod_{p<q}E_{\ell_p-\ell_q}^{\rho_{pq}}\right)v_0.\label{eq:gsv2Na}\eeq

\bigskip

\nt{\bf Example: N=1}\\
We have: ~$ r_1=\sigma_1=2h_1-2r_1$. It follows that ~$
\sigma_1=r_1=(2h_1+1)/3$~ and consequently, cf. \cite{ADDS}, \beq
\label{n3} v_s=c \left( \sum_{a=1}^{3} G_a^2 -2\Lambda(M) K
\right)^{p/2} \sum_{\sigma=0,\sigma_1}E_{\ell_1}^{\sigma}\,v_0 \eeq

\bigskip

\nt{\bf Example: N=2}\\
We have: $i=1, j=2$ and $\sigma_1=0, \sigma_2=2h_2-2r_2+1$\\
Under the conditions, which must be satisfied, it follows that
$\lambda_{12}=0$ and $\rho_{12}=r_2-r_1+h_1-h_2+1$. On the other
hand $r_1=\rho_{12},\quad r_2=\sigma_2-\rho_{12}$. This provide \beq
\rho_{12}=(6h_1-h_2+3)/5,\quad \sigma_2=2(2h_1+h_2+1)/5\nn\eeq
Thus, the general form of the singular vector for ~$n=2N+1=5$~ is:
\beq\label{n5} v_s=c \left( \sum_{a=1}^{5} G_a^2 -2\Lambda(M) K
\right)^{p/2} \sum_{\sigma=0,\sigma_2\,;\
\rho=0,\rho_{12}}E_{\ell_2}^{\sigma}E_{\ell_1-\ell_2}^{\rho}\,v_0
\eeq

\subsubsection{Invariant equations}

Substituting the right actions  into the singular vectors
\eqref{eq:gsv2Na}  and performing similar computation to $ n=2N$
case, we obtain the invariant equations \beq \left(
\sum_{a=1}^{2N+1} \frac{\partial^2}{\partial x_a^2} + 2M \del{t}
\right )^{p/2}
\sum_{\sigma,\lambda,\rho}\left(\pi_R(E_{\ell_N})\right)^\sigma
\left(\prod_{r<s}\pi_R(E_{\ell_r+\ell_s})^{\lambda_{rs}}\prod_{p<q}\pi_R(
 E_{\ell_p-\ell_q})^{\rho_{pq}}\right)\psi =
 \psi' \label{hier2na}\eeq where $\pi_r(E_{\ell_N}),\pi_R(E_{\ell_r+\ell_s}), \pi_R(E_{\ell_p-\ell_q})$
 are given in \cite{ADDS}, ~$ m = -\Lambda(M) $, and ~$\psi'$ belongs to the
representation ~$\L'$~ such that ~$\L'(M)=\L(M)$,
~$\L'(H_k)=\L(H_k)+r_k$ and ~$\L'(D)=\L(D)+p = \D+p = \ha (p+2N+3)$.

In particular for $ N=1$ one obtains \cite{ADDS}, cf. \eqref{n3}:
\beq
 \left( \sum_{a=1}^{3} \frac{\partial^2}{\partial x_a^2} + 2M \del{t} \right)^{p/2}
 \sum_{\sigma=0,\sigma_{1}}\left(\del{\varphi_1}\right)^\sigma\psi=\psi'
 \label{hier2.11}, \eeq while for $N=2$ the invariant equation is obtained from \eqref{n5}:
\beq \left( \sum_{a=1}^{5} \frac{\partial^2}{\partial x_a^2} + 2M
\del{t}
 \right)^{p/2}
 \sum_{\sigma,\rho}\left(\del{\varphi_2}+\eta_{12}(\del{\varphi_1}+\del{\xi_{12}})\right)^\sigma
 \left(\del{\eta_{12}}+\eta_{23}\del{\eta_{13}}\right)^\rho\psi~=
 ~\psi'
 \label{hier2.21}  \eeq
where $\sigma=0,\sigma_2\,;\ \rho=0,\rho_{12}\,$.

\np

\setcounter{equation}{0}
\section{Non-relativistic invariant equations for ~$\hs(3)$}

\subsection{Algebraic structure and actions}

We consider the case $n=3$ (which is the $N=1$ odd case of the previous
section) separately since it is most relevant for the physical applications.

 {\bf Triangular decomposition:}  $\sch{3} =
\sch{3}^+\oplus\sch{3}^0\oplus\sch{3}^-$ \bea
\sch{3}^+ & = & \{G_0, G_+, G_-, K, E_+\}\nonumber\\
\sch{3}^0 & = & \{D, M, H\}\nonumber\\
\sch{3}^- & = & \{P_0, P_+, P_-, P_t,
E_-\}\label{eq:cartandec}.\eea Here we have simplified the notations
as follows: \bea E_+ & = &
E_{l_1}=-\frac{1}{\sqrt{2}}(J_{13}-\II J_{23})\nonumber\\
E_- & = & E_{-l_1}=-\frac{1}{\sqrt{2}}(J_{13}+\II J_{23})\;\;,\;\;H_1=H= -J_{12}\nonumber\\
G_0 & = & \sqrt{2}G_3\;\;,\;\;G_+=G_1+\II G_2\;\;,\;\; G_-=G_1-\II
G_2\nonumber\\
P_0 & = & \sqrt{2}P_3\;\;,\;\;P_+=P_1+\II P_2\;\;,\;\; P_-=P_1-\II
P_2\label{eq:gcomplex}.\eea We recall that this complexification
reflects on the coordinates which read \beq x_0=
\frac{1}{\sqrt{2}}x_3\;\,\;\;x_+=x_1-\II x_2\;\;,\;\;x_-=x_1+\II x_2.\eeq\\

Taking into account (\ref{sch}, \ref{eq:gcomplex}) the non-vanishing
commutation relations are:\bea && [P_t,D] =
2P_t\;\;,\;\;[P_{0,\pm},D]=P_{0,\pm}\;\;,\;\;[P_t,G_{0,\pm}]=P_{0,\pm
}\nonumber\\ && [ P_t,K ] =
D\;\;,\;\;[P_{0,\pm},K]=G_{0,\pm}\;\;,\;\;
[P_0,G_0]=[P_{\pm},G_{\mp}]=2M \nonumber\\
&& [D,G_{0,\pm }] = G_{0,\pm }\;\;,\;\;[D,K]=2K \nonumber\\
&& [P_0, E_{\pm}]= -P_{\mp}\;\;,\;\;[P_{\pm},
E_{\pm}]=P_0\;\;,\;\;[P_{\pm}, H]=\pm P_{\pm}\nonumber\\
&& [G_0, E_{\pm}]= -G_{\mp}\;\;,\;\;[G_{\pm},
E_{\pm}]=G_0\;\;,\;\;[G_{\pm}, H]=\pm G_{\pm}\nonumber\\
&& [E_{\pm},H]=\mp E_{\pm}\;\;,\;\;[E_+, E_-]=-H.\eea\\

{\bf Gauss decomposition}: Let $g\in \hat{S}$ is the element of the
Schr\"odinger group, than its triangular decomposition is $g =
g_+g_0g_-$, where\bea g_+ & = &
e^{x_0G_0}e^{x_+G_+}e^{x_-G_-}e^{tK}e^{\varphi_+E_+}\nonumber\\
g_0 & = & e^{\delta D}e^{mM}e^{hH}\nonumber\\
g_- & = &
e^{v_0P_0}e^{v_+P_+}e^{v_-P_-}e^{kP_t}e^{\varphi_-E_-}\label{eq:gdec}.\eea

\bigskip
\noindent{\bf Remark:} The explicit expressions for $g_+, g_0,
g_-$ can be easily obtained from the following relations valid in the matrix representation (\ref{Repsn})
(in terms of $8\times 8 $ matrices):
\bea
M^q & = & 0, ~q> 1 \nonumber\\
D^{2q+1} & = & D\;\;,\;\;H^{2q+1}  = H \ , ~q\geq 0\label{eq:Dexp}\\
D^{2q} &=& D^2 \;\;,\;\;H^{2q}=H^2\ , ~q >0\label{eq:Hexp}\\
G_0^q  & = &  G_+^q=G_-^q=G_0G_+=G_0G_-=G_+G_-=0, ~q> 1\label{eq:Gexp}\\
P_0^q & = & P_+^q=P_-^q=0\;,\;q> 1\nn\\ P_t^q & = & 0\;,\; ~q>
1\nn\\
K^q & = & 0\;,\; ~q> 1\nn\\
E_+^q & = & 0 \;\;,\;\; ~q>2\nn\\
E_-^q & = & 0 \;\;,\;\; ~q>2\label{eq:eminusexp},\eea where in
the above expressions $q$ is an integer number. Furthermore the
matrix realization of the algebra $\sch{3}$ may also used  be
obtain the left(right) action.

\bigskip

{\bf  Left action:}
\begin{itemize}\item
 Action of $\sch{3}^0$. Using the commutation relations and matrix realization of
 $\sch{3}$, up to terms quadratic in $\tau $ it
 follows
 \bea && e^{-\tau M}g_+=g_+e^{-\tau M}\nonumber\\
&& e^{-\tau D}g_+=\prod_{a=1}^3\exp\left(x_ae^{-\tau}
G_a\right)\exp\left(te^{-2\tau}K\right) e^{\varphi_+E_+}e^{-\tau
D}\nonumber\\
&& e^{-\tau H}g_+=e^{x_0G_0}e^{x_+e^\tau
G_+}e^{x_-e^{-\tau}G_-}e^{\varphi_+e^{-\tau}E_+}e^{-\tau
H}\label{eq:gzeroact}.\eea The above expressions provide the
following representation by the left action:\bea \pi_L(M) & = &
-\Lambda(M)\nonumber\\
\pi_L(D)& = & -\Lambda(D)-\sum_{a=1}^3x_a\frac{\partial}{\partial
x_a}-2t\frac{\partial}{\partial t}\nonumber\\
\pi_L(H) & = & -\Lambda(H)+x_+\frac{\partial}{\partial
x_+}-x_-\frac{\partial}{\partial
x_-}-\varphi_+\frac{\partial}{\partial
\varphi_+}\label{eq:g0left}.\eea \item Action of $\sch{3}^+$
\bea e^{-\tau G_0}g_+ & = & e^{(x_0-\tau)G_0}e^{x_+G_+}e^{x_-G_-}e^{tK}e^{\varphi_+E_+}\nonumber\\
 e^{-\tau
G_+}g_+ & = & e^{x_0G_0}e^{(x_+-\tau )G_+}e^{x_-G_-}e^{tK}e^{\varphi_+E_+}\nonumber\\
e^{-\tau G_-}g_+ & = & e^{x_0G_0}e^{x_+G_+}e^{(x_--\tau
)G_-}e^{tK}e^{\varphi_+E_+}\nonumber\\ e^{-\tau K}g_+ & = &
e^{x_0G_0}e^{x_+G_+}e^{x_-G_-}e^{(t-\tau)K}e^{\varphi_+E_+}\nonumber\\
e^{-\tau E_+}g_+ & = & e^{(x_0+\tau x_+)G_0}e^{x_+G_+}e^{(x_--\tau
x_0)G_-}e^{tK} e^{(\varphi_+-\tau)E_+}\label{eq:gplusact}\eea
provide\bea \pi_L(G_0) & = & -\frac{\partial}{\partial x_0},\qquad
\pi_L(G_+) =-\frac{\partial}{\partial x_+},\qquad \pi_L(G_-) =
-\frac{\partial}{\partial x_-}\nonumber\\\pi_L(K) & = &
-\frac{\partial}{\partial t},\qquad \pi_L(E_+)=
x_+\frac{\partial}{\partial x_0}-x_0\frac{\partial}{\partial
x_-}-\frac{\partial}{\partial \varphi_+}\label{eq:gplusleft}\eea
\item
 Action of $\sch{3}^-$~:
\bea
\pi_L(P_0) & = & -\Lambda(M)x_0-t\frac{\partial}{\partial x_0}\nonumber\\
\pi_L(P_+) & = & -\Lambda(M)x_+-t\frac{\partial}{\partial x_+} \nonumber\\
\pi_L(P_-) & = & -\Lambda(M)x_--t\frac{\partial}{\partial x_-}\nonumber\\
\pi_L(P_t) & = & -\Lambda(D)t-\half \Lambda(M)(x_0^2+x_+^2+x_-^2)
-t\sum_{a=1}^3x_a\frac{\partial}{\partial
x_a}-t^2\frac{\partial}{\partial t}\nonumber\\
\pi_L(E_-) & = & -\Lambda(H)\varphi_++x_-\frac{\partial}{\partial
x_0}-x_0\frac{\partial}{\partial
x_+}-\half\varphi_+^2\frac{\partial}{\partial\varphi_+}\label{eq:schminusrep}\eea
\end{itemize}

\bigskip

{\bf Right action:}
\begin{itemize}
\item By definition\beq
\pi_R(M)=\Lambda(M)\;\;,\;\;\pi_R(D)=\Lambda(D)\;\;,\;\;\pi_R(H)=\Lambda(H)\eeq
\item The right action of $\sch{3}^+$ is given by (again up to
terms quadratic in $\tau$): \bea g_+e^{\tau G_0} & = &
e^{(x_0+\tau)G_0}e^{x_+G_+}e^{(x_-+\tau\varphi_+)G_-}e^{tK}e^{\varphi_+E_+}\nonumber\\
g_+e^{\tau G_+} & = &
e^{(x_0-\tau\varphi_+)G_0}e^{x_+G_+}e^{(x_-\tau\half\varphi_+^2)G_-}
e^{tK}e^{\varphi_+E_+}\nonumber\\
g_+e^{\tau G_-} & = &
e^{x_0G_0}e^{x_+G_+}e^{(x_-+\tau)G_-}e^{tK}e^{\varphi_+E_+}\nonumber\\
g_+e^{\tau K} & = & e^{x_0G_0}e^{x_+G_+}e^{x_-G_-}e^{(t+\tau
K)}e^{\varphi_+E_+}\nonumber\\ g_+e^{\tau E_+} & = &
e^{x_0G_0}e^{x_+G_+}e^{x_-G_-}e^{tK}e^{(\varphi_++\tau)E_+},\label{eq:gplusract}\eea
which provides\bea \pi_R(G_0) & = & \frac{\partial}{\partial
x_0}+\varphi_+\frac{\partial}{\partial x_-}\nonumber\\
\pi_R(G_+) & = & \frac{\partial}{\partial
x_+}-\varphi_+\frac{\partial}{\partial
x_0}-\half\varphi^2_+\frac{\partial}{\partial x_-}\nonumber\\
\pi_R(G_-) & = & \frac{\partial}{\partial
x_-}\;\;,\;\;\pi_R(K)=\frac{\partial}{\partial
t}\;\;,\;\;\pi_R(E_+)=\frac{\partial}{\partial
\varphi_+}\label{eq:right}.\eea
\end{itemize}

\subsection{Singular vectors}

 In this case a basis in the Verma module
$V^{\Lambda}\cong U(\sch{3}^+)v_0$ is \beq\label{eq:basis}
v_{\omega\al\be\gamma\sigma}=G_0^\omega G_+^\al G_-^\be K^\gamma
E_+^\sigma v_0,\eeq where $\omega, \al, \be, \gamma, \sigma $ are
nonnegative integers. Using the properties of lowest
weight vector \bea Mv_0 & = & \Lambda(M)v_0=Mv_0\nonumber\\
Dv_0 & = & \Lambda(D)v_0=\D v_0\nonumber\\
Hv_0 & = & \Lambda(H)v_0=-hv_0,\eea we first calculate the action
of the Cartan generators on the basis (\ref{eq:basis})\bea
Mv_{\omega\al\be\gamma\sigma} & = &
\Lambda(M)v_{\omega\al\be\gamma\sigma}\nonumber\\
Dv_{\omega\al\be\gamma\sigma} & = &
(p+\D)v_{\omega\al\be\gamma\sigma}\;\;,\;\;p=\omega + \al +\be
+2\gamma\nonumber\\
Hv_{\omega\al\be\gamma\sigma} & = &
(r-h)v_{\omega\al\be\gamma\sigma}\;\;,\;\;r=-\al +\be+\sigma .
\label{eq:cartan}\eea In the same manner we calculate the action
of generators from $\sch{3}^-$\bea\label{lact}
P_0v_{\omega\al\be\gamma\sigma} & = & 2\omega\Lambda(M)v_{\omega -
1, \al\be\gamma\sigma}+\gamma v_{\omega +1, \al\be,\gamma-1,
\sigma}\\
P_+v_{\omega\al\be\gamma\sigma} & = & 2\be\Lambda(M)v_{\omega ,
\al,\be -1, \gamma\sigma}+\gamma v_{\omega,\al+1,\be,\gamma-1,
\sigma} \nn\\
P_-v_{\omega\al\be\gamma\sigma} & = & 2\al\Lambda(M)v_{\omega ,
\al-1,\be\gamma\sigma}+\gamma v_{\omega,\al,\be+1,\gamma-1,
\sigma} \nn\\
P_tv_{\omega\al\be\gamma\sigma} & = &
\omega(\omega-1)\Lambda(M)v_{\omega-1,\al,\be,
\gamma\sigma}+2\al\be\Lambda(M)v_{\omega,\al-1,\be-1,\gamma,
\sigma }\ +\nonumber\\
 && +  \gamma(p-\gamma+\D-1)v_{\omega\al\be,\gamma-1,\sigma}\nn\\
E_-v_{\omega\al\be\gamma\sigma} & = & \omega v_{\omega-1
,\al+1,\be\gamma\sigma}-\be v_{\omega+1,\al,\be-1,\gamma,
\sigma}-\half\sigma(\sigma-1+2h)v_{\omega\al\be\gamma\sigma-1} \nn \eea
Next we apply the requirement $v_{\omega\al\be\gamma\sigma}$ to be
homogeneous, from which follows that $r$ and $p$  are to be fixed. Thus,
  from five variables $\omega, \al, \be, \gamma,\sigma$
only three are independent. We choose $\omega, \al, \sigma $ as
independent variables, which means that the singular vectors must
have the form\beq \label{eg:sivegform} v_s=\sum_{\omega,\al,
\sigma}a_{\omega\al\sigma}v_{\omega\al\be\gamma\sigma} ,\eeq
the sum obeying the following:\\
 The replacement $\al \to \al\pm 1$
causes the change of dependent variables $$ \be \to \be\pm
1\;\;,\;\;\gamma\to\gamma\mp 1 $$   The
replacement $\omega \to \omega \pm 2$ causes the change
$\gamma\to\gamma\mp 1$ and do not reflect on the others.

Next we use the definition of a singular vector\beq H v_s =
\Lambda'(H) v_s\;\;,\;\;H\in \sch{3}^0\;\;;\;\;
Xv_s=0\;_;,\;\;X\in\sch{3}^-\label{eq:defsvecz},\eeq to obtain
\begin{enumerate}\item $P_0v_s=0$ \bea &&
2\omega\Lambda(M)a_{\omega\al\sigma}v_{\omega-1,\al\be\sigma}+\gamma
a_{\omega\al\sigma} v_{\omega +1, \al\be,\gamma-1,
\sigma}\nonumber=\\
&& = \left(2\omega\Lambda(M)a_{\omega\al\sigma}+(\gamma+1)
a_{\omega-2,\al\sigma}\right) v_{\omega -1,
\al\be\gamma\sigma}=0\nonumber\\
 &&
a_{\omega\al\sigma}=-\frac{\gamma+1}{2\Lambda(M)\omega}a_{\omega-2,
\al\sigma}\label{eq:cpzero}\eea \item $P_+v_s=0$\beq
a_{\omega\al\sigma}=-\frac{\gamma+1}{2\Lambda(M)\be}a_{\omega,
\al-1, \sigma}\label{eq:cpplus}\eeq \item $P_-v_s=0$\beq
a_{\omega\al\sigma}=-\frac{\gamma+1}{2\Lambda(M)\al}a_{\omega,
\al-1, \sigma}\label{eq:cpminus}\eeq

\item $P_tv_s=0$\eqn{ptvs}
(\omega+1)(\omega+2)\Lambda(M)a_{\omega+2,\al\sigma}+2(\al+1)^2\Lambda(M)a_{\omega,\al+11,\be-1,\sigma
}+\gamma(p-\gamma+\D-1)a_{\omega\al\sigma}=0\nonumber\eeq
 \item $E_-v_s=0$ \beq (\omega
a_{\omega,\al-1,\sigma}-\al a_{\omega-2,\al, \sigma}) v_{\omega-1
,\al, \be-1,
\gamma+1,\sigma}+\half\sigma(\sigma-1-2h)a_{\omega\al\sigma}v_{\omega\al\be\gamma\sigma-1}=0\label{eq:newcondition},\eeq
\end{enumerate}
We note that from (\ref{eq:cpzero}) the substitution $\omega
=1$ leads to $a_{\omega\al\sigma}=0$ for $\omega$ odd, while from
(\ref{eq:cpplus}, \ref{eq:cpminus}) follows that
$\al=\be$. Further substitution of (\ref{eq:cpzero}, \ref{eq:cpplus}) in \eqref{ptvs}
gives\bea && \gamma(p/2+\D-5/2)=0\nonumber\\
&& p=\omega+2\al+2\gamma=(2\D+5)/2\label{eq:cptime}.\eea Then $p$
must be also even.

Solving the above recurrence relations was done in \cite{ADDS}. The result
for the most general form of the singular vector is: \beq\label{eq:vectsing}
v^{p,\sigma }_s=c\left(\half
G_0^2+G_+G_--2\Lambda(M)K\right)^{p/2}E_+^{\sigma}v_0,\eeq where ~$\sigma$~ has two values:
$\sigma=0,1+2h\in\bbn$,  $p=5-2\D\in 2\bbn$,
$c$ arbitrary nonzero constant.

Furthermore, one can verify that $ v^\sigma_s=E_+^{\sigma}v_0 = v^{0,\sigma}_s$
is also a singular vector. Really the relations \beq
P_0v^\sigma_s=P_+v^\sigma_s=P_-v^\sigma_s=P_tv^\sigma_s=0\eeq are
automatically satisfied. The last condition\beq
E_-v^\sigma_s=\half\sigma(\sigma-1-2h)v^\sigma_s=0\eeq is
satisfied again for $\sigma=1+2h=1-2\Lambda(H)$.

Thus, ~$v^{p,\sigma }_s$~ is a composite singular vector and
 we have a quartet to Verma modules with weights ~$\L = (\L(M)=M,\L(D)=\D,\L(H)=-h)$,
~$\L_p= (M,p+\D,-h)$,
~$\L'_\sigma= (M,\D,\sigma-h)$,
~$\L_{p,\sigma}= (M,p+\D,\sigma-h)$, which can be given in an embedding commutative diagram:
\eqnn{diagv} V^\L & \longrightarrow & V^{\L_p} \cr
&&\cr
\downarrow && \downarrow \cr
&&\cr
V^{\L'_\sigma} & \longrightarrow & V^{\L_{p,\sigma}}
\eea
where each arrow points ~$to$~ the embedded module.
By construction, ~$V^\L$~ has singular vectors ~$v^{p,0}_s\,$~ and ~$v^{0,\sigma}_s\,$,
~$V^{\L'_\sigma}$~ has singular vector ~$v^{p,0}_s\,$,
~$V^{\L_p}$~ has singular vector ~$v^{0,\sigma}_s\,$.

All above properties follow from the following relations: \bea &&
Mv^{p,\sigma}_s=\Lambda(M)v^{p,\sigma}_s \nonumber\\
&&
Dv^{p,\sigma}_s=(\Lambda(D)+p)v^{p,\sigma}_s \nonumber\\
&& Hv^{p,\sigma}_s=(\Lambda(H)+\sigma )v^{p,\sigma
}_s \label{eq:repsplit}\eea

\subsection{Non-relativistic equations}

The formula (\ref{eq:vectsing})
provides (taking the right action) the following form of invariant
differential operator and correspondingly invariant equation \bea
&& {\cal D}_{p,\sigma} =
\left(\sum_{i=1}^3\frac{\partial^2}{\partial
x_i^2}+2M\del{t}\right)^{p/2}\left(\del{\varphi_+}\right)^\sigma \nonumber\\
&& {\cal D}_{p,\sigma}\,\psi(t,x_1,x_2,x_3;\varphi_+)=\psi'(t,x_1,x_2,x_3;\varphi_+)\ .
\label{eq:splitoper}\eea
where ~$\psi$~ belongs to the representation ~$C^\L$~ characterized by ~$\Lambda$ , while
~$\psi'$~ belongs to the representation ~$C^{\L_{p,\sigma}}$~ characterized by ~$\Lambda_{p,\sigma}$.

Of course, analogously to the Verma module embedding picture \eqref{diagv} there is a quartet
commutative diagram:
\eqnn{diageq} C^\L & \longrightarrow & C^{\L_p} \cr
&&\cr
\downarrow && \downarrow \cr
&&\cr
C^{\L'_\sigma} & \longrightarrow & C^{\L_{p,\sigma}}
\eea
where each vertical  arrow depicts the differential operator ~${\cal D}_{0,\sigma} ~=~ \left(\del{\varphi_+}\right)^\sigma$,
while each horizontal arrow depicts the differential operator ~${\cal D}_{p,0}\,$.

\np
\setcounter{equation}{0}
\section{$q$-Schr\"odinger algebra}

\subsection{$q$-deformation of the \S}

In this Section we review \cite{DDMq}.
For other approaches to $q$-deformations of the \S\ we refer to
\cite{CarowWatamura,BCGST,FlVi,Ubriaco,TerAntonian,Nersessian,BHP,BHPa,BHPb,BHPc,BHNN,BHNNa,BHNNb,Aizq,Aizqa,Carroll,LWC,Chakraborty,Gerasimov,AoNiSu,KawYos,KMY}.

We use the following  $q$-number notations:
\eqn{qq}
[a]_q ~\doteq~ {{ q^{a} - q^{-a} \over q - q^{-1} }} ~, \qquad
[a]'_q ~\doteq~ [a]_{q^{1/2}} ~=~
{{ q^{a/2} - q^{-a/2} \over q^{1/2} - q^{-1/2} }} ~=~
{[a/2]_q \over [1/2]_q }
\eeq
and similarly for any diagonal operator  instead of ~$a$.

Now we construct a
$q$-deformation ~$\hs_q (1) $ of the \S{}
under the following  conditions:
\begin{itemize}
\item{1.} A realization of the generators ~$P_t, P, G, K$ in terms
          of $q$-difference operators and multiplication operators should be
          available.
\item{2.} In the limit $ q \to 1 $ we should have the classical relations
           \eqref{scha}.
\item{3.} The subalgebra structure should be preserved by the deformation and
          especially the d-deformed $sl(2,\bbc)$ subalgebra
          generated by $D,K,P_t$ should coincide
          with the usual Drinfeld-Jimbo deformation ~$U_q(sl(2,\bbc))$.
\end{itemize}

\nt
With these conditions we get
for $\hs_q (1) $
the following nontrivial relations instead of \eqref{scha}~:
\eqna{schq}
&&\hpt\ \hg\ -\ q\ \hg\ \hpt ~=~ \hx \\  &
&[\hx, \hk]  ~=~ \hg ~q^{-D}\\ &
&[D, \hg] ~=~ \hg\\ &
&[D, \hx] ~=~ -\hx\\ &
&[D, \hpt] ~=~ -2\hpt\\ &
&[D, \hk] ~=~ 2\hk\\ &
&[\hpt, \hk] ~=~ [ D ]_q\\  &
&\hx\ \hg\ -\ q^{-1}\ \hg\ \hx ~=~ m \\ &
&\hpt\ \hx\ -\ q^{-1}\ \hx\ \hpt   ~=~ 0\eena

Conditions 2.\ and 3.\ can now be checked directly, \eqreff{schq}{e,f,g} are
the standard commutation relations of the Drinfeld-Jimbo deformation
~$U_q(sl(2,\bbc))$. Moreover we obtain a $q$-deformed centrally extended
Galilei subalgebra generated by $P_t,P,G$. The deformation is a "mild" one,
in the sense
that commutators are turned into $q$-commutators, cf. \eqreff{schq}{a,h,i} and
it differs from the Galilei algebra $q$-deformation given in \cite{BCGST},
which is not a surprise taking into account that the latter
is not a subalgebra of a ($q$-deformed) \S. Condition 1.\ will be discussed
later.

The commutation relations \eqref{schq}  are graded  as
the undeformed ones \eqref{scha} if we define the grading as in
\eqref{graa}.

For ~{\it real}~ $q$~ the commutation relations \eqref{schq} are preserved also by
 the involutive  antiautomorphism \eqref{coj} supplemented by the
 condition ~$\om (q) ~=~ q$.
With this conjugation the subalgebra generated by ~$P_t+K,i(P_t-K),D$,
(the fixed points of $\om$), is ~$U_q(sl(2,\bbr))$.

\subsection{Lowest weight modules of ~$\hs_q(1)$}

We consider  lowest weight modules (LWM) of $\hs_q(1)$,
in particular, Verma modules, in complete analogy with the
undeformed case $q=1$. In particular, the lowest weight vector fulfils
\eqref{bact}.
 The Verma module ~$V^\D$~ is given explicitly by
~$V^\D ~=~ U_q(\cs^+)~\otimes ~v_0\,$,  where
~$U_q(\cs^+)$~ is  the $q$-deformed  universal
enveloping algebra of $\cs^+\equiv   \cs(1)^+$.
Clearly, ~$U_q(\cs^+)$~  has the basis
elements ~$p_{k,\ell} ~=~ G^k K^\ell$.
The basis vectors of the Verma module are
~$v_{k,\ell} ~=~ p_{k,\ell} \otimes v_0$,
(with $v_{0,0} = v_0$). The action of the
$q$-\S\ on this basis is derived easily from \eqref{schq}:
\eqna{vactq}
&
&D~v_{k,\ell} ~~=~~ (k + 2\ell +\D)~ v_{k,\ell}\\ &
&G~v_{k,\ell} ~~=~~ v_{k+1,\ell}\\ &
&K~v_{k,\ell} ~~=~~ v_{k,\ell+1}\\ &
P_x v_{k\ell} ~~=&~~ q^{1-k\over 2} \
M \, [k]'_q ~ v_{k-1,\ell} ~+~
q^{1-\D-\ell-k} \ [\ell]_q ~ v_{k+1,\ell-1}\\
&
P_t ~ v_{k\ell}  ~~=&~~
[\ell]_q ~ [k+\ell-1+\D]_q ~ v_{k,\ell-1} ~+
~ M ~ {[k]'_q ~
[k-1]'_q \over [2]'_q}
~ v_{k-2,\ell}\eena
For the derivation of \eqref{vactq}
the following relations (which follow from  \eqref{schq})
are usefull:
\eqna{vct} &
&P_xG^k ~-~ q^{-k} G^k P_x ~=~ M\ q^{(1-k)/2}
[k]'_q ~G^{k-1}\\   &
&P_xK^\ell ~-~ K^\ell P_x ~=~ q^{1-\ell}
[\ell]_q ~G ~K^{\ell-1} ~q^{-D}\\ &
&P_t G^k ~-~ q^k G^k P_t ~=~ [k]_q ~ G^{k-1} ~P_x
~+~ {[k]'_q ~[k-1]'_q \over [2]'_q} ~G^{k-2}\\
&
&P_tK^\ell ~-~ K^\ell P_t ~=~
[\ell]_q ~ K^{\ell-1}~ [D+\ell-1]_q\eena

Because of \eqref{vactq}{a} we notice that the Verma module $V^\D$
can be decomposed in homogeneous subspaces
w.r.t. grading operator $D$ as in \eqref{vdec}.

Next we analyze the reducibility of ~$V^\D$~ through
    singular vectors. The considerations are exactly similar to
the undeformed case.
All possible singular vectors were given explicitly  in
\cite{DDMq} as follows.
Fix the grade ~$p>0$~ and denote
the singular vector as ~$v^p_s$~. Consider the
case of ~{\it even}~ grade, ~$p\in 2\bbn$. Since ~$v^p_s \in V^\D_p$~
we have:
\eqn{sngq} v^p_s ~~=~~ \sum_{\ell =0}^{p/2}
~a_\ell ~v_{p-2\ell,\ell} ~=~ \cq^p(G,K)
\ \otimes \ v_0
~, \quad p ~{\rm even} \eeq
Applying \eqref{sing} we obtain that a singular vector exists only for
$\D = {3-p \over 2}$ (as for $q=1$) and is given explicitly
for arbitrary ~$q$~ by the formula:
\eqnn{sngaq}
v^p_s ~=&~ a_0 ~ \sum_{\ell=0}^{{p \over 2}} ~ \left( ~ - ~ M
            [2]'_q ~ \right)^\ell {{p \over 2}
\choose \ell}_q ~ v_{p-2\ell,\ell} ~=
~a_0 ~\left( G^2 ~-~ M ~[2]'_q ~K \right)^{p \over 2}_q
\ \otimes \ v_0 \cr
&\cq^p(G,K) ~=~ a_0 ~\left( G^2 ~-~ M ~[2]'_q ~K \right)^{p \over 2}_q \eea
where
\eqn{nwb}
{p \choose s}_q ~\doteq ~ {[p]_q! \over  [s]_q! [p-s]_q!} \,, \qquad
[n]_q! ~\doteq ~ [n]_q  [n-1]_q \ldots [1]_q \eeq
For ~{\it odd}~ grade there are no singular vectors as for $q=1$.

To analyze the
consequences of the reducibility of
our Verma modules
we take the subspace of ~$V^{(3-p)/2}$~:
\eqn{sbsq}I^{(3-p)/2} ~~=~~ U_q(\cs^+) ~v^p_s \eeq
It is invariant under the action of the $q$-deformed \S,
and is
isomorphic to a Verma module ~$V^{\D'}$ with shifted weight
~$\D' = \D+p = (p+3)/2$. The latter Verma module has no singular
vectors.

Let us denote by ~$\cl^{(3-p)/2}$ the factor--module
~$V^{(3-p)/2} /I^{(3-p)/2}$~ and by ~$\vr p\rg$~ the lowest weight
vector of $\cl^{(3-p)/2}$.
As a consequence of \eqref{sing} and \eqref{sngaq}
~$\vr p\rg$~ satisfies:
\eqna{facq}&
&P_x ~\vr p\rg ~~=~~ 0\\  &
&P_t ~\vr p\rg ~~=~~ 0\\  &
&\sum_{\ell=0}^{{p \over 2}} ~ \left( ~ - ~ m ~
            [2]'_q ~ \right)^\ell
{{p \over 2}
\choose \ell}_q ~ G^{p-2\ell}~ K^\ell
~\vr p\rg ~~=~~ 0\eena

Now from \eqreff{facq}{c} we see that:
\eqn{exprq} K^{p/2}  ~\vr p\rg ~~=~~ -
\sum_{\ell=0}^{p/2-1} ~{ 1
\over \left(- m [2]'_q  \right)^{p/2-\ell} }
~{p/2 \choose \ell}_q ~
G^{p-2\ell} ~ K^\ell ~\vr p\rg \eeq
By a repeated application of
this relation to the basis one can get rid of all
powers ~$\geq p/2$~ of ~$K$.
Thus the basis of ~$\cl^{(3-p)/2}$~
will be a ~{\it singleton basis}~ for ~$p=2$, and a
~{\it quasi--singleton basis}~ for $p\geq 4$~:
\eqn{degbq} {\rm dim}~V^{(3-p)/2}_n ~~=~~ 1 ~, \quad
{\rm for}~ n=0,1 ~{\rm or} ~n\geq p \eeq
and it is given by:
\eqn{basiq} v^p_{k\ell} ~ \equiv ~
G^k ~K^\ell ~\vr p\rg ~, \qquad p\in 2\bbn, ~
k,\ell \in\bbz_+\,, ~\ell \leq p/2 -1, ~\D = {3-p\over 2} \eeq

The transformation rules of this basis are
\eqref{vactq} except \eqreff{vactq}{c} for ~$\ell ~=~ p/2 -1$, when we have:
 $$ K~v^p_{k,p/2-1} ~~=~~ -
\sum_{s=0}^{p/2-1} ~{ 1
\over \left(- m [2]'_q  \right)^{p/2-s} }
~{p/2 \choose s}_q ~v^p_{k+p-2s,s} \eqno(\reff{vactq}{c'}) $$
{}From the transformation rules we see that
~$\cl^{(3-p)/2}$~ is irreducible.
In the simplest case ~$p=2$~ the irrep ~$\cl^{1/2}$~ is
also an irrep of the $q$-deformed
centrally extended Galilean subalgebra ~$\hhg_q(1)$~
generated by ~$P_x,P_t,G$.

Hence, the complete list of the irreducible lowest weight
modules over the $q$-deformed
centrally extended \S\ is given by \cite{DDMq}:
\begin{itemize}
\item{} ~$V^\D$~, ~when ~$d\neq (3-p)/2$, ~$p\in 2\bbn$ ;
\item{} ~$\cl^{(3-p)/2}$, ~when ~$\D= (3-p)/2$,
~$p\in 2\bbn$ .
\end{itemize}
These irreps are infinite-dimensional.

\vskip 5mm

\subsection{Vector--field realization of ~$\hs_q(1)$~ and
generalized $q$-deformed
heat equations}

\nt
Let us introduce the "number" operator ~$N_y$~
for the coordinate ~$y=x,t$, i.e.,
\eqn{nyd}
N_y ~y^k ~ = ~ k ~y^k, \eeq
and the
~$q$ - difference operators ~$\cd_y\,$, $\cd'_y\,$,
which admit a general
definition on a larger domain than polynomials, but on
polynomials are well defined as follows:
\eqna{doq}&
&\cd_y ~\doteq~ {1 \over y } [ N_y]_q\\   &
&\cd'_y ~\doteq~ {1 \over y [\half]_q } \left[{ N_y \over 2} \right]_q
           ~ = ~ {1 \over y} [N_y]_q',  \eena
so that for any suitable function $f$ we obtain as a consequence of \eqref{nyd} :
\eqna{nyf} &
&\cd_y ~ f(y) ~ = ~ {f(qy)~-~f(q^{-1}y) \over y ~ (q-q^{-1})}\\ &
&\cd_y' ~ f(y) ~ = ~ {f(q^{\half}y) ~ - ~ f(q^{-\half}y) \over y
(q^{\half}-q^{-\half})}\eena
For ~$q ~\to ~1$~ one has: ~$N_y ~\to ~y\pd_y\,$,
~$\cd_y,\, \cd'_y ~\to \pd_y\,$.

With this notation there exists \cite{DDMq}
a five-parameter realization of \eqref{schq} via $q$-difference operators
(or vector--field realization for short):
\eqna{qfx}&
&P_t ~ = ~ q^{c_1} ~ \cd_t ~ q^{(1-c_5)N_t+(1-c_4)N_x}\\ &
&P_x ~ = ~ q^{c_2} ~ \cd_x' ~ q^{-c_4N_t+(c_3+\half)N_x}\\ &
&D ~ = ~ 2N_t ~ + ~ N_x ~ + ~ \D\\  &
& G ~ = ~ q^{c_2-c_1-c_4+c_5} ~ t ~ \cd_x' ~
q^{(c_5-c_4)N_t +(c_3+c_4-\half)N_x} \cr
&&+ ~ q^{-c_2-c_3-\half} ~ mx ~ q^{c_4N_t-(c_3+1)N_x} \\ &
&
K ~ = ~ q^{-c_1+c_5-1-\D} ~ t^2 ~ \cd_t ~ q^{(c_5-1)N_t + c_4 N_x} \cr
&&+ ~ q^{-c_1+c_5-1-\D} ~ tx ~ \cd_x ~ q^{(c_5-2)N_t+(c_4-1)N_x} \cr
&&+ ~ q^{-c_1+c_5-1} ~ [\D]_q ~ t ~ q^{c_5N_t+c_4N_x} \cr
&&+ ~ q^{-2c_2-3c_3-\trhalf-\D} ~ [\half]_q ~ mx^2 ~
q^{2(c_4-1)N_t-2(c_3+1)N_x}  \eena
where ~$c_1,c_2,c_3,c_4,c_5$~ are arbitrary parameters.
(There might be other
vector--field realizations that are not equivalent to
the one just given.)

For $q=1$ we recover the standard
vector--field realization of ~$\hs(1)$ \eqref{vecb}.

Our realization \eqref{qfx} may be used to construct a polynomial realization
of the irreducible lowest weight modules considered in
Section 3. For that case we represent
the lowest weight vector by the function 1.
Indeed, the constants in \eqref{qfx}
are chosen so that \eqref{bact}\ is satisfied:
\eqn{bacfq} D~1 ~~=~~ \D ~, \quad P_x~1 ~~=~~ 0,
\quad P_t~1 ~~=~~ 0 \eeq
Applying the basis elements ~ $p_{k,\ell} = G^k K^\ell$~
of the universal enveloping  algebra $U_q(\cs^+)$ to 1
we get polynomials in ~$x,t$ which
will be denoted by ~$f_{k,\ell} ~\equiv ~ p_{k,\ell}~1$.
For the explicit expressions we refer to \cite{DDMq}.
        There it was also shown   that
the basis ~$f_{k,\ell}$~ is a realization
of the irreducible lowest weight representations of ~$\hs_q(1)$~
listed at the end of the previous section. ~Indeed,
there is 1-to-1 correspondence between the
states ~$v_{k,\ell}$~ of the Verma modules over ~$\hs_q(1)$~
and the polynomials ~$f_{k,\ell}\,$. The irreducible
lowest weight representations of ~$\hs_q(1)$~ are
factor--modules of Verma modules, with factorization
over the invariant subspaces generated by
singular vectors.
This statement
is trivial if
there is no singular vector.
When a singular vector exists,
i.e., for the representations $V^{(3-p)/2}$, we
 first
obtain a $q$-difference operator
by substituting in $\cq^p(G,K)$ (cf. \eqref{sngq}, \eqref{sngaq})
each generator with its vector--field realization.
For the irreducibility of $\cl^{(3-p)/2}$ it is enough to
show that
the $q$-difference operator $\cq^p(G,K)$
vanishes identically when applied to 1.
This contains more information as $\cq^p(G,K)$
gives also a $q$-difference equation invariant under
the action of $\hs_q (1)$. Because of this invariance
the solutions of this equation are elements of
$\cl^{(3-p)/2}$.
Thus we have
an infinite family of $q$-difference equations,
the family members being labelled by $p\in 2\bbn$, i.e., we have
one equation for each representation space $V^{(3-p)/2}$.
These equations may be  called  generalized $q$-deformed heat equations
($m$ real) or generalized $q$-deformed Schr\"odinger equations
($m$ imaginary).
The case ~$p=2$~ is a $q$-difference analog of
the ordinary heat/Schr\"odinger equation.

Before making the last example explicit we make a choice of constants
in \eqref{qfx}
and set for simplicity
~$c_1=c_2=c_3=c_4=c_5=0$~
 so that to work with
 simpler expressions for the generators:
\eqna{qfs}
P_t ~ &=& ~ \cd_t ~ q^{N_t+N_x}\\
P_x ~ &=& ~ \cd_x' ~ q^{\half N_x}\\
D ~ &=& ~ 2N_t ~ + ~ N_x ~ + ~ \D\\
G ~ &=& ~ t ~ \cd_x' ~ q^{-\half N_x} ~
+ ~ q^{-\half} ~ M x ~ q^{-N_x}\\
 K ~ &=& ~ q^{-\D-1} ~ t^2 ~ \cd_t ~ q^{- N_t}
~ + ~ q^{-\D-1} ~ tx ~ \cd_x ~ q^{-2 N_t - N_x}\ + \nn\\
&&+ ~ q^{-1} ~ [\D]_q ~ t ~
~ + ~ q^{-\D-\trhalf} ~ [\half]_q ~ M x^2 ~ q^{-2 N_t- 2 N_x} \eena

The operator ~$ S_q ~ =~ {\cal Q} ~=~ G^2 ~ - ~ [2]'_q M ~ K $~
determining the singular vectors reads:
\eqnn{qht}
S_q ~ &= & ~ t^2 ~ q^{\half}~  \left(~
\cd_x'^2 ~ q^{-N_x} ~- ~ q^{-\D-\trhalf} ~
[2]_q' ~M ~ \cd_t ~ q^{-N_t} ~ \right) ~+ \nn\\
&& ~ +~ M t x~ \cd_x' ~
\left( ~ [2]'_q ~-~ \left( 1 + q^{N_x}\right)
q^{-\D-1- 2N_t} ~ \right)
~ q^{-\trhalf N_x} ~+\nn\\
&& ~ + q^{-1} ~ M t ~ \left( ~ q^{-2N_x} ~-~ [2\D]'_q  ~ \right) ~+ \nn\\
&& ~ + q^{-2} ~ M^2 x^2 ~ \left( ~ 1- q^{-\D+\half -2N_t} ~ \right)
~ q^{-2N_x} \eea
which for $ q=1 $ gives:
\eqn{schrop} S ~ = ~ t^2 ~ ( ~ \partial_x^2 ~ - ~ 2M ~ \partial_t ~ ) ~
          + ~M ~ t ~ (1-2\D) \eeq
Hence we interpret for ~$ \D = \half $~
(which corresponds to the lowest singular vector $(p=2)$)
the equation $ S_q f = 0 $ as a
~{\it $q$-deformed heat/Schr\"odinger equation}~
as we motivated in the Introduction.
The explicit form of this
equation is:
\eqnn{qsva}
 S_q ~ f ~ &=& ~ 0 \cr &&\cr
S_q ~ &=& ~ t^2 ~ q^{\half}~  \left(~
\cd_x'^2 ~ q^{-N_x} ~- ~ q^{-2} ~
[2]_q' ~M ~ \cd_t ~ q^{-N_t} ~ \right) ~+ \cr
&& ~ +~ M tx~ \cd_x' ~
\left( ~ [2]'_q ~-~ \left( 1 + q^{N_x}\right)
q^{-\trhalf- 2N_t} ~ \right)
~ q^{-\trhalf N_x} ~- \cr
&& ~ - ~\l q^{-1} ~ M tx ~ \cd_x ~ q^{-N_x}
~+   ~\l q^{-2} ~ M^2 x^2 t ~\cd_t ~ q^{-N_t-2N_x} \eea
where ~$\l ~\doteq ~ q - q^{-1}\,$.

 This is the proposal of \cite{DDMq} for a ~{\it q-deformed heat equation}.
For ~$q\mt 1$~ ($\l \mt 0$) this equation
leads to
the ordinary heat/Schr\"odinger equation.

\vskip 5mm

{\bf Remark:} We note that there exists another $q$--deformation of the vector--field
realization of $\hs(1)$ given by Floreanini and Vinet \cite{FlVi}.
They start with a special $q$-deformed heat equation
and look  for a $q$-symmetry algebra on its solution variety.
The resulting $q$-deformation of the \S\  in \cite{FlVi},
which we  call ~{\it on shell}~ deformation
is different from  the one of \cite{DDMq} and
is   valid only on the solutions of the
$q$-deformed heat equation under consideration.~$\diamondsuit$

\np

\setcounter{equation}{0}
\section{Difference analogues of the free Schr\"odinger equation}

\subsection{Motivations}

In this Section we review \cite{DDMd}.
For other approaches to difference equations with \S\ symmetry we refer to
\cite{FlVi,LVW,Kozlov,Barker}.

The time evolution of physical systems is generically described via
(partial) differential equations, especially via the \sgl\ for the
case of non-relativistic quantum mechanics. The use of such
equations, however, is an idealization, because the infinitesimal
structure inherent in the
definition of differential operators,
\beq \partial_x \, f(x_0) ~ \equiv
 ~ \lim_{\zeta \to 0} \, {f(x_0 + \zeta) \, - \,
f(x_0-\zeta) \over 2\zeta}\ , \eeq
cannot be reproduced in physical measurements.
A {\em realistic} measurement of a "differential quantity"
such as velocity etc.\ actually involves measurements at two {\em distinct}
points in the physical space-time, i.e., it is based on measurements
at $x$ and $x+\zeta$ with $\zeta$ {\em finite} and not infinitesimal.
Hence, in a realistic physical setting, only finite difference
quotients of non-infinitesimal quantities should occur.
The physical space-time could be either continuous or discrete for
the space and/or time coordinates.
On very small length scales (Planck scale)
it is likely
that some ``grained'' or ``lattice'' structure is more appropriate as
an ``arena'' for physical theories than a continuous space-time.

Apart from such more fundamental considerations there are also practical
reasons for the use of finite difference equations.
Generically, one
has to use {\em approximations} in order to get a quantitative description of
a physical system, and sometimes lattice models are useful in this
context. In these models the objects of the theory are allowed to
``live'' only on discrete points of a lattice. It is obvious that in
this setting  finite difference operators involving the lattice
spacing are basic quantities.

There are various types of (finite) difference operators.
The usual choice, used also in the context of lattices, are
operators of additive type:
\beq
\label{fin-difx}
D_x \, f(x) ~ = ~ {f(x+\zeta) \, - \, f(x-\zeta) \over 2\zeta}\ ,
\eeq
i.e., functions at ~$x \pm \z$~ are compared.
(Note, that $\lim_{\zeta \to 0} D_z = \partial_z$.)
Another type which was recently
being discussed in the context of generalized symmetries are difference
operators of multiplicative type, the so called $q$-difference operators:
\beq
\label{fin-difq}
D^q_x \, f(x) ~ = ~ {f(qx) \, - \, f(q^{-1}x) \over x(q-q^{-1})}
\eeq
which compare functions at ~$qx$ and $q^{-1}x\,$. They appear naturally in the
representation theory of quantum groups or more generally for $q$-deformed symmetries
as we discussed in the previous Section.

If one wants to model physical systems through difference operators one has
to derive or to motivate, e.g., evolution equations as difference equations.
A formal attempt uses of a kind of correspondence  principle. This means to
replace the usual differential operator by a difference operator so that
in the continuum limit the "usual" theory is reproduced.

A more generic method is the following: If a differential equation or
relation is derived from first principles, e.g., from a symmetry group or
an algebra, one can try to formulate already the principles  in terms of
difference operators. If necessary, one has to change the derivation
and possibly also
further assumptions are needed. Certainly also here we will get in the
limiting case a differential equation, however, the result may differ
from the one obtained from the correspondence principle.

Here  we discuss the free quantum mechanical
Schr\"odinger equation (SE) without spin on
~$\bbr^n_x \times \bbr_t$~  based on first principles:
The SE is physically characterized through
representation theory of the central extension of the $(n + 1)$-dimensional
\salg\ \sae.
We gave in previous sections a purely algebraic construction for the family of \sae\
invariant {Schr\"odinger} equations from a family of
singular vectors in Verma modules over \sae. (We
used also $q$-difference operators to derive a $q$-analogue of the
Schr\"odinger equation, cf.\ \cite{DDMq} and the previous Section.)
Now, we use the general method and in the realization
of \sae\ through vector fields we replace the vector fields
with additive difference vector fields, i.e., vector fields with difference
operators instead of differential operators. Because the construction of
Verma modules and the construction of invariant differential equation is
completely algebraic we can apply this method also in this case.

To relate the \salg\ invariance as a first principle for a quantum
mechanical evolution equation with difference operators we need the
definitions and some properties of these operators and  a
realization of \sae\ through difference vector fields. This construction is
not unique: additional assumptions which are physically well motivated
are necessary. The essential point in our derivation is the observation
made in \cite{DDM} (and in previous Sections) that the construction of
invariant equations from Verma modules is independent of the
realization of the generators of the corresponding algebra. The reason for
this is that the construction of
\cite{DDM} is completely algebraic, i.e., uses only the commutation
relations of the algebra.

\subsection{Definition and notation}

As we discussed above the natural candidate for the formulation of
difference analogues of differential equations are
additive finite difference operators as in (\ref{fin-difx}).
In the representation of \sae\ one needs (partial) finite difference
operators with respect to space-coordinates $x_i$ and the
time-coordinate $t$. We use $\tau$
as ``fundamental (time-)length'' for the time-coordinate and $\xi$ as
"fundamental (space-) length" for the space coordinate, i.e.
\beq
\begin{array}{l@{}l}
D_t \, f(t,x) ~ &\equiv ~ {f(t+\tau,x) \, - \, f(t-\tau,x) \over
2\tau},
\\ \\
D_x \, f(t,x) ~ &\equiv ~ {f(t,x+\xi) \, - \, f(t,x-\xi) \over
2\xi}.
\end{array}
\eeq
The generalization to the $n+1$-dimensional case is obvious.
Note that \eqref{fin-difx} is not the only possibility for the
introduction of a finite difference operator with the ``correct''
limit; one could, e.g., use any expression of the
form:\footnote{The requirement $a,b \in \bbz$ reflects the fact that only
entire multiples of the fundamental length $\zeta$ are considered
measurable.}
$$ D_z^{(a,b)} ~ \doteq ~ {f(z+a\zeta) \, - \, f(z-b\zeta) \over (a+b)\zeta}
\qquad a,b \in \bbz,~ \quad a \neq -b. $$
Thus, $D_z = D_z^{(1,1)}$.

In the context of finite difference operators {\bf shift operators}
$T_z{}^a$, defined as
\beq
\label{fin-shift}
T_z{}^a \, f(z) ~ \equiv ~ f(z+a\zeta), \qquad a \in \bbz
\eeq
are useful.
We have
$$ D_z^{(a,b)} ~ = ~ {T_z^a \, - \, T_z^{-b} \over (a+b) \zeta}. $$
Apart form the symmetric difference operator \eqref{fin-difx}, we will also
use a "forward difference operator"
\beq
\label{fd-forw}
D_z^+ ~ \equiv ~ D_z \, T_z{} ~ = ~ {T_z{}^2 \, - \, 1 \over 2\zeta}
\eeq
and a "backward difference operator"
\beq
\label{fd-bw}
D_z^- ~ \equiv ~ D_z \, T_z{}^{-1} ~ = ~ {1 \, - \, T_z{}^{-2} \over
2\zeta}.\eeq
For our purposes it is important that $D_z^{(a,b)}$ and $T_z^a$ are,
by definition, {\em linear operators}.

\subsection{Construction of a realization of \sae}
We construct a realization of of \sae\ with finite difference and
shift operators. As explained above this is possible
only if one imposes additional {\em assumptions} which we choose to be
the following:
\begin{sl}
\begin{enumerate}
\label{fd-ass}
\item In the zero-limit of the ``fundamental lengths'' the representation
      \eqref{veca} should be recovered.
\item The number of additional terms with vanishing limit in the
      representation should be ``as small as possible''.
\item The generators of space translations
      $P_i$ (and of time translations $P_t$ when we consider representations
      with difference operators in $t$) should take a ``simple form'',
      i.e.,\ they should be realized by terms:\footnote{We use
      the shorthand notation $T_i \equiv T_{x_i}$ and $D_i \equiv D_{x_i}$.}
      $$ P_i ~ = ~ D_i \, T_i^a, \quad a \in \bbz. $$
\end{enumerate}
\end{sl}

It turns out that the following operators constitute a realization of
\sae\ obeying the above mentioned assumptions, cf.\ \cite{DDMd}
\eqnn{fd-real}
P_t & = &  D_t^+, \\[3ex]
P_i & = & D_i^+, \nn\\[3ex]
D & = &  2t \, D_t^+ ~ + ~ \sum_{i=1}^n \, x_i \, D_i^- ~ + ~ \D, \nn\\[3ex]
J_{ij} & = & x_j \, D_i^+ \, T_j^{-2} ~ - ~
                   x_i \, D_j^+ \, T_i^{-2} \nn\\[3ex]
G_i & = &  t \, D_i^+ \, T_t{}^{-2}~ + ~ M \, x_i \, T_i^{-2}, \nn\\[3ex]
K & = & (t^2 - 2 \tau t) \, D_t^- \, T_t^{-2}~ + ~
            t\, \left(\sum_{i=1}^n
\, x_i \, D_i^- ~\right) \, T_t{}^{-2} ~ \nn\\[1ex]
& & ~+ ~
 {M \over 2} \sum_{i=1}^n \, \left( x_i \, T_i^{-2} \right)^2
       ~ + ~ t\D \, T_t{}^{-2} \nn\eea
By setting $\tau \to 0$, i.e.
$$ D_t^\pm \to \partial_t, \qquad T_t^a \to 1 $$
of $\xi \to 0$, i.e
$$ D_i^\pm \to \partial_i, \qquad T_i^a \to 1 $$
one obtains realizations of \sae\ in which only space
(respectively time)
differentials are replaced by difference operators.
We stress the fact that all these realizations of \sae\ are {\em linear}.

\subsection{Invariant finite difference equations}

Above we obtained \sae -invariant partial
differential equations by inserting the \sae -realization \eqref{veca}
into the expression \eqref{sngz} which determines the singular
vectors of the corresponding \verma s. We mentioned already
that the validity of this result does {\em not} depend on whether one
has a realization with {\em differential} operators or not. In fact
any {\em linear} realization leads to invariant equations. Thus, we
can insert \eqref{fd-real} into \eqref{sngz} and obtain \sae
-invariant finite difference equations. As a result, we find that the
equations
\beq
\left(D_t^+ ~ - ~ {1 \over 2M} \,
\left(\sum_{i=1}^n \, D_i^+\right)^2\right)^{p \over 2} \;
\psi (t,x) ~ = ~ 0, \quad p ~ {\rm even}.
\eeq
are invariant under the \sae -realization
\eqref{fd-real} with $\D={n+2-p \over 2}. $

As a special case ($p=2$) we
obtain a finite difference analogue of the free \sgl\ in $n$
space dimensions:
\eqn{fd2-sgl1}
\left(D_t^+ ~ - ~ {1 \over 2M} \,
\left(\sum_{i=1}^n \, D_i^+\right)^2\right) \;
\psi(t,x) ~ = ~ 0.\eeq

Setting $\tau \to 0$ or $\xi \to 0$ in \eqref{fd-real}, respectively,
we obtain discrete-continuous analogues of the free
\sgl in which only space- or time- differentiation is replaced
with the corresponding difference operators, i.e.
\eqn{pd1}
\left(\partial_t ~ - ~ {1 \over 2M} \, \left(\sum_{i=1}^n \, D_i^+\right)^2
\right) \;
\psi(t,x) ~ = ~ 0.\eeq
or respectively
\eqn{pd2}
\left(D_t^+ ~ - ~ {1 \over 2M} \, \left(\sum_{i=1}^n \,
\partial_i\right)^2\right) \;
\psi(t,x) ~ = ~ 0.\eeq

We stress that all these analogues of the free \sgl\ are {\em not}
postulated but derived by our algebraic construction.
Especially the appearance of $D^+$ -- instead of another $D^{(a,b)}$ or
linear combinations of several such operators -- is forced by the
assumptions given above and the construction of the equations.

In the case $n=1$ equation (\ref{fd2-sgl1}) was considered in
\cite{FlVi,LVW}, equations (\ref{pd1}),(\ref{pd2}) -
in \cite{LVW}.
These authors looked for the symmetry of these equations employing
difference or differential-difference operators,
and they found that on the solution set of the equations these
operators satisfy the $n=1$ \salg, though the
explicit expressions of the operators are different
from ours (and for (\ref{fd2-sgl1}) between the two papers mentioned).

\np

\section*{Acknowledgments}
The author cordially thanks Prof. Phua Kok Khoo    for the kind invitation to write this review. ~
The author cordially thanks his coauthors of reviewed papers ~N. Aizawa, H.-D. Doebner, C. Mrugalla, S. Stoimenov,
for their input   and  for discussions. ~The author
has received partial support from COST action MP-1210.

\end{document}